\documentclass[aps,pra,twocolumn,superscriptaddress]{revtex4-2}
\usepackage[utf8]{inputenc}
\usepackage{soul}
\usepackage{textcomp}
\usepackage[english]{babel}
\usepackage{graphicx}% Include figure files
\usepackage{dcolumn}% Align table columns on decimal point
\usepackage{bm}% bold math
\usepackage{bm}
\usepackage{amsmath}
\usepackage{verbatim}     %needed for comments
\usepackage[dvipsnames]{xcolor}       %needed for comments
\usepackage[normalem]{ulem}   %needed for strikethrough font
\usepackage{bbold}
\usepackage{mathrsfs}
\usepackage[mathscr]{eucal}
\usepackage{physics}
\newcommand{\beq}{\begin{equation}}
\newcommand{\eeq}{\end{equation}}
\newcommand{\bei}{\begin{itemize}}			% Begin itemize environment %
\newcommand{\eei}{\end{itemize}}			% End itemize environment %

\usepackage{hyperref}
\hypersetup{
   pdfpagemode=None,
   pdfstartpage=1,
   pdfmenubar=true,
   pdftoolbar=true,
   colorlinks = true,
   linkcolor=blue,
   citecolor=blue,
   urlcolor=blue,
   bookmarksopen=false
 }

\begin{document}
\title{Ultra-long quantum walks via spin-orbit photonics}
\author{Francesco Di Colandrea}\email{francesco.dicolandrea@unina.it}
\affiliation{Dipartimento di Fisica, Universit\`{a} degli Studi di Napoli Federico II, Complesso Universitario di Monte Sant'Angelo, Via Cintia, 80126 Napoli, Italy}
\author{Amin Babazadeh}
\affiliation{Dipartimento di Fisica, Universit\`{a} degli Studi di Napoli Federico II, Complesso Universitario di Monte Sant'Angelo, Via Cintia, 80126 Napoli, Italy}
\author {Alexandre Dauphin}\email{alexandre.dauphin@icfo.eu}
\affiliation{ICFO -- Institut de Ciencies Fotoniques, The Barcelona Institute of Science and Technology, 08860 Castelldefels (Barcelona), Spain}
\author {Pietro Massignan}
\affiliation{Departament de F\'isica, Universitat Polit\`ecnica de Catalunya, Campus Nord B4-B5, 08034 Barcelona, Spain}
\author{Lorenzo Marrucci}
\affiliation{Dipartimento di Fisica, Universit\`{a} degli Studi di Napoli Federico II, Complesso Universitario di Monte Sant'Angelo, Via Cintia, 80126 Napoli, Italy}
\affiliation{CNR-ISASI, Institute of Applied Science and Intelligent Systems, Via Campi Flegrei 34, 80078 Pozzuoli (NA), Italy}
\author{Filippo Cardano}\email{filippo.cardano2@unina.it}
\affiliation{Dipartimento di Fisica, Universit\`{a} degli Studi di Napoli Federico II, Complesso Universitario di Monte Sant'Angelo, Via Cintia, 80126 Napoli, Italy}

\begin{abstract}
The possibility of fine-tuning the couplings between optical modes is a key requirement in photonic circuits for quantum simulations. In these architectures, emulating the long-time evolution of particles across large lattices requires sophisticated setups, that are often intrinsically lossy. Here we report ultra-long photonic quantum walks across several hundred optical modes, obtained by propagating a light beam through very few closely-stacked liquid-crystal metasurfaces. By exploiting spin-orbit effects, these implement space-dependent polarization transformations that mix circularly polarized optical modes carrying quantized transverse momentum. As each metasurface implements long-range couplings between distant modes, by using only a few of them we simulate quantum walks up to 320 discrete steps without any optical amplification, far beyond state-of-the-art experiments. To showcase the potential of this method, we experimentally demonstrate that in the long-time limit a quantum walk affected by dynamical disorder generates maximal entanglement between two system partitions. Our platform grants experimental access to large-scale unitary evolutions while keeping optical losses at a minimum, thereby paving the way to massive multi-photon multi-mode quantum simulations.

\end{abstract}

\maketitle

The demand for versatile and efficient machines implementing quantum operations %simulators of quantum dynamics 
in a controlled and accessible environment \cite{Buluta2009} has fueled the development of a variety of artificial systems, based for instance on ultra-cold atomic ensembles \cite{Gross2017}, trapped ions \cite{Blatt2012}, superconducting \cite{Kjaergaard2020} and photonic platforms \cite{Wang2020}. These devices promise access to revolutionary applications in quantum computing \cite{Zhong2020} and quantum information processing \cite{Slussarenko2019,Bogaerts2020,Blais2020}, and allow for the generation of highly-entangled states, to be used for example in quantum metrological tasks \cite{Pezze2018}. 
Among quantum processes that are investigated within these systems, evolutions known as quantum walks (QWs) \cite{Venegas2012} stand out for their applications in fields as diverse as quantum computation \cite{Gong2021}, transport phenomena \cite{Mares2020} and topological physics \cite{Kitagawa2010}. QWs model the quantum evolution of particles (or ``walkers") equipped with a spin-like internal degree of freedom (the ``coin"), moving on complex graphs. In their simplest version, discrete-time QWs involve a walker moving along a one-dimensional (1D) lattice, whose sites are labeled by an integer number $m$. After $\tau$ discrete steps, a QW maps the input state $\ket{\psi(0)}$ to $\ket{\psi(\tau)}=U^\tau\ket{\psi(0)}$, where $U$ is the single-step evolution operator. In photonic simulations of QWs, as in those of generic tight-binding Hamiltonians, position states $\ket{m}$ are mapped into distinct optical modes, which are coupled so as to engineer the desired evolution \cite{Schreiber2012,Broo10_PRL,Cardano2017,Defienne2016,Tang2018b,Wang2020}. These photonic circuits, providing a linear and unitary map between a set of input and output modes, are the core-technology at the basis of boson sampling experiments \cite{Brod2019} that recently brought to the demonstration of quantum advantage with photonic setups \cite{Zhong2020,Madsen2022}. While optical circuits typically rely on the manipulation of spatial modes arranged in separated optical paths, either in integrated \cite{Wang2020,Arrazola2021} or in bulk interferometers \cite{Zhong2020}, the control of co-propagating optical modes having complex temporal \cite{Schreiber2012,Madsen2022} or spatial \cite{DErrico2020,Plachta2022,Goel2022} structures is emerging as a powerful technology to generate reconfigurable and high-dimensional photonic machines, not to mention their obvious applications in the generation and manipulation of structured light \cite{Forbes2021}. 
In this work, we present a platform implementing unitary operations on co-propagating optical modes having circular polarization and carrying quantized transverse momentum. By leveraging state-of-the-art spin-orbit photonics \cite{Cardano2015b}, that enables the control of light spatial structure through the manipulation of its polarization, we engineer optical evolutions that are equivalent to QWs of hundreds of steps and involve a massive amount of modes. Significantly, this is achieved by using only three spin-orbit optical metasurfaces \cite{Rubano2019}. As such, this setup turns out to be extremely compact and efficient, and overcomes crucial limitations of standard photonic circuits simulating QWs and other quantum evolutions.

\begin{figure*}[t]
\includegraphics[width=1\linewidth]{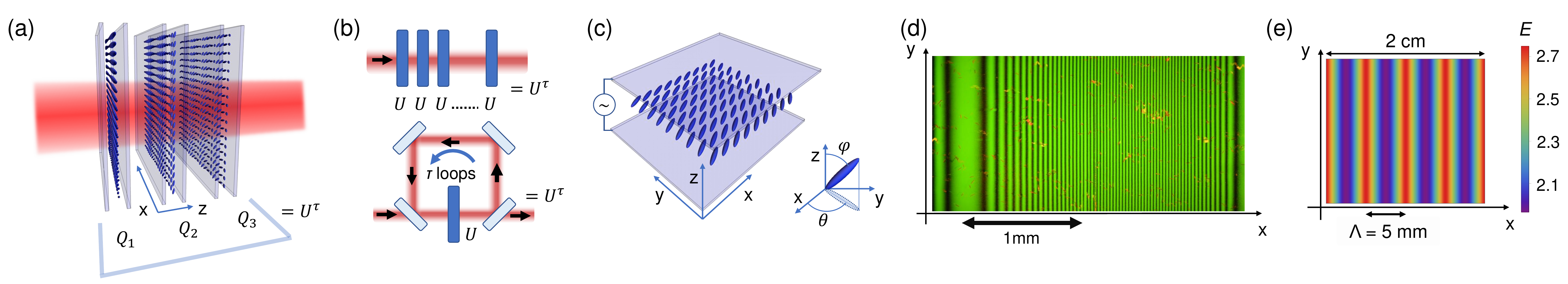}
\caption{{\bf Implementing QWs  via liquid-crystal metasurfaces.} (a) In our devices, a light beam passes through three liquid-crystal metasurfaces (LCMSs), whose combination implements the evolution operator $U^\tau$  of the entire walk comprising $\tau$ steps. (b) In usual QW setups, instead, the number of optical operations scales linearly with the number of steps. (c) Scheme of a LCMS. 
The out-of-plane orientation $\varphi$ of the LC molecules can be easily tuned by an external electric field. (d) A LCMS, fabricated to realize 240 steps of a QW, is observed between crossed polarizers, unveiling its birefringent structure. (e) We plot a map in the $xy$ plane of the argument $E$ of the eigenvalues associated with the step operator obtained from the LCMS reported in panel (d). The profile is clearly periodic with spatial period $\Lambda$. %Our plates actually extend for 2cm (hosting 4 BZs).
}
\label{fig:panels}
\end{figure*}
\section*{Results}
\noindent \textbf{Optical modes with quantized transverse momentum.} 
The photonic circuit we present here couples circularly polarized optical modes carrying quantized transverse momentum, thus displaying a linear phase gradient in a direction that is perpendicular to the main optical axis of the system. By labelling the optical axis as $z$ and the quantization direction as $x$, these modes have the following expression:  
%%%%
\begin{equation}
\ket{m,j}=A(x,y,z)e^{ik_z z}e^{i k_m x}\ket{j}.
\label{eqn:modes}
\end{equation}
%%%%
They carry a quantized amount of transverse momentum $k_m=m \Delta k_\perp$, where $\Delta k_\perp=2\pi/\Lambda$,
$m$ is an integer number, and $\Lambda$ is the spatial period of the transverse beam profile. In Eq.~\eqref{eqn:modes}, $A(x,y,z)$ is the spatial profile of the beam and $k_z$ is the wavevector $z$-component, while $\ket{j}$ ($j\in\{L,R\}$) represents the polarization state, being left-circular or right-circular, respectively. These photonic states represent Gaussian beams that propagate in the $xz$ plane and form a small angle $\alpha\simeq m\lambda/\Lambda$ with the $z$ axis. As detailed in Ref.\ \cite{DErrico2020}, they are orthogonal %(not overlapping) 
as long as $w_0\gtrsim\Lambda$, with $w_0$ being the beam radius at the waist. In our setup, all these modes propagate within a single optical beam, and their ensemble may therefore be regarded as a discrete photonic synthetic dimension \cite{Celi2014,Yuan2018}.

From the perspective of a QW, a photon being in the state $\ket{m,j}$ represents a walker sitting at the lattice site $m$, with a coin state $\ket j$. QWs involving these modes have been first reported in Ref.\ \cite{DErrico2020} (up to 5 steps in a 2D QW) and in Ref.\ \cite{DErrico2021} (up to 14 steps in a 1D QW), by exploiting light propagation through a sequence of polarization gratings ($g$-plates), whose number was proportional to that of simulated steps. Here we construct the overall photonic transformation, that maps input and output states via a QW evolution, by means of only three spin-orbit metasurfaces, as shown in Fig.\ \ref{fig:panels}(a). This represents a radically different approach from standard architectures (as for instance those in Refs.\ \cite{DErrico2020,DErrico2021}) relying on cascaded or looped setups, where the number of passages through optical plates increases linearly with the number of steps $\tau$ (see Fig.\ \ref{fig:panels}(b)). 

%In our setup, ultra-long evolutions are achieved without any optical amplification, which is essential instead in recent demonstrations of similar QWs \cite{Weidemann2022}. 

Specifically, we implement a photonic simulator consisting of a collimated light beam - or equivalently single photons - propagating along the $z$ axis through suitable optical metasurfaces. The encoding of the walker position $m$ into the photon transverse momentum $k_m$ automatically maps the walker momentum $q$ into the photon transverse position $x$. % being these conjugate variables in their respective domain. 
 The Hilbert space associated with the walker momentum (from now on the $x$ variable) provides the convenient framework to model translation-invariant evolutions. In this case, indeed, Bloch theorem dictates that the operator $U$ is composed of a family of block-diagonal unitary operators $\mathcal U(x)$ that act on the coin only. 
As the two orthogonal coin states are mapped into left and right circular polarizations $\ket L$ and $\ket R$, the Bloch operator $\mathcal{U}(x)$ can be effectively realized through a position-dependent polarization rotation, as those widely used to shape multi-mode structured light \cite{Piccardo2022}. Such optical transformation is necessarily periodic, with a spatial period $\Lambda$ that defines the effective length of the synthetic %Brilluoin
Brillouin zone (BZ) for the quasi-momentum. By manipulating the coin state at each quasi-momentum value $x$, we simulate the dynamics of a particle hopping between discrete states $\ket{m,j}$.\\ 

%%%%
\begin{figure*}[!t]
\includegraphics[width=\linewidth]{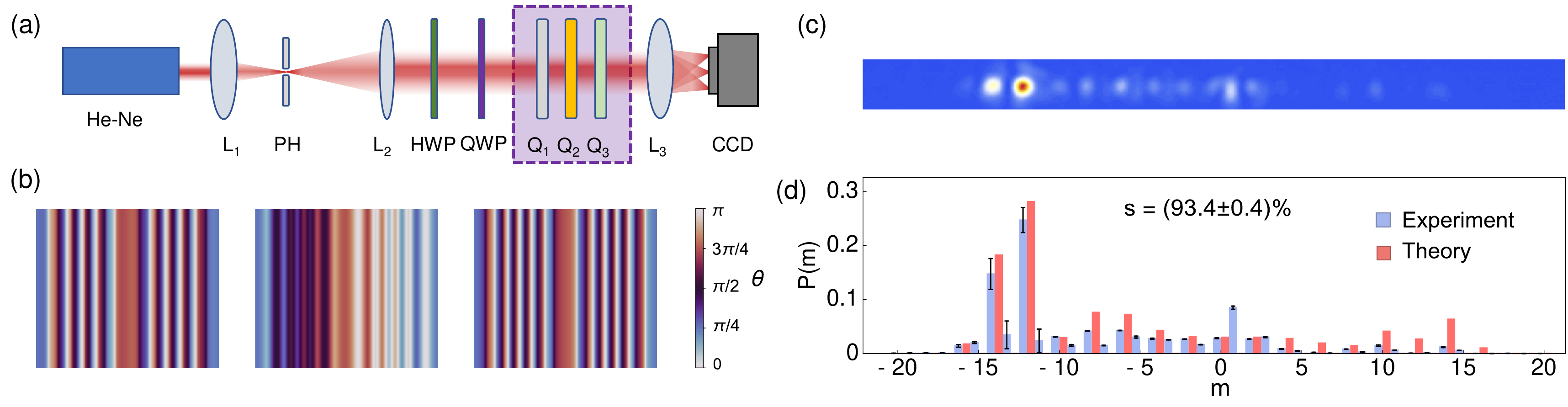}
\caption{{\bf Generating QW evolutions.}
(a) Sketch of the experimental setup. After preparing the input state, %the laser beam passes through the three LCMSs generating the whole evolution. 
the whole evolution is generated by the three LCMSs (Q$_1$, Q$_2$, Q$_3$).
The final distribution is %acquired by 
recorded by a camera placed in the focal plane of a Fourier lens. 
(b) Optic-axis modulations of the three LCMSs implementing $U_1^{20}(\pi)$. The plots cover one BZ. %(c) Reduced Stokes parameters $s_x$ (top) and $s_y$ (bottom) of the light emerging from the QW setup, for a left-circular polarized input. Measured values are compared to numerical results.
(c) %Full 
Light intensity distribution recorded for one $U_1^{20}(\pi)$ realization. 
(d) Experimental distribution resulting from $U_1^{20}(\pi)$, compared with the theoretical prediction. }
\label{fig:distributions}
\end{figure*}
%%%

\noindent \textbf{Spin-orbit photonics for QWs.} To generate a 1D QW with a
two-state coin (such as those discussed in Ref.\ \cite{Cardano2017}) performing $\tau$ discrete steps, we first compute its complete Bloch operator $\mathcal{U}^\tau (x) $. Then, we implement the associated polarization transformation using patterned liquid-crystal (LC) waveplates, hereafter referred to as liquid-crystal metasurfaces (LCMSs). These are manufactured by sandwiching a micro-metric LC layer between two glass plates, whose internal sides are coated with a transparent conductive material. The latter allows one to apply an electric field perpendicular to the LC slab, tuning in turn the effective birefringence of the cell and its associated optical retardation $\delta$ (see Fig.\ \ref{fig:panels}(c)). 
In the basis of circular polarizations, the action of a single LCMS is provided by the following matrix: 
%%%%
\begin{equation}
Q_\delta(\theta)=
\begin{pmatrix}
\cos\left(\frac{\delta}{2} \right) & i\sin\left(\frac{\delta}{2} \right)e^{-2i\theta}\\
i\sin\left(\frac{\delta}{2} \right)e^{2i\theta} & \cos\left(\frac{\delta}{2} \right)

\end{pmatrix}
,
\label{eqn:waveplate}
\end{equation}
%%%%
where $\theta$ is the angle that LC molecular direction forms with the $x$ axis. We consider a set of three plates, having fixed optical retardations $\delta_1=\delta_3=\pi/2$ and $\delta_2=\pi$. The corresponding cells are thus acting as quarter-wave plates (QWP) and half-wave plates (HWP), respectively. By adjusting the value of $\theta$ of individual devices, one can implement an arbitrary polarization rotation via the minimal sequence $\mathcal L=Q_{\pi/2}(\theta_3)Q_{\pi}(\theta_2)Q_{\pi/2}(\theta_1)$ \cite{Simon1990}. The latter cannot be further reduced to two elements, as the unitary matrix $\mathcal U^\tau$ has three independent real parameters (neglecting a global phase factor). LC technology at the basis of standard $q$-plates \cite{Rubano2019} enables us to realize metasurfaces with a continuously-modulated angular orientation $\theta(x)$, which is key to implementing the desired spatially-varying polarization transformation $\mathcal L(x)=\mathcal{U}^{\tau}(x)$. Naturally, the same functionality could be obtained by using spin-orbit metasurfaces based on a different technology, such as dielectric nanostructures \cite{Devlin2017,Devlin2017b} or other metamaterials \cite{Dorrah2022}. 

In Fig.\ \ref{fig:panels}(d) we show an experimental image of a prototypical LCMS fabricated for the experiment, observed between crossed polarizers. The action of the crossed polarizers maps different values of $\theta(x)$ into levels of transmitted light, thus converting the angular pattern into an intensity distribution. Our QW is then realized by cascading three LCMSs, as shown in Fig.\ \ref{fig:panels}(a), replacing long sequences of optical devices that are normally required for these simulations. As an example, our previous setup \cite{DErrico2020} for simulating QWs required two optical elements at each time-step, introducing nearest-neighbour couplings only and thus limiting our experiments to at most 14 time-steps \cite{DErrico2021}. Here instead each metasurface couples hundreds of modes with a long-range connectivity, dramatically reducing the number of required elements.
Actual values of the patterns $(\theta_1(x), \theta_2(x), \theta_3(x))$ implementing the desired operator $\mathcal U^\tau$ are found via analytical expressions, that are illustrated in the Methods. The overall polarization transformation is described by a position-dependent SU(2) operator that exhibits a periodic character (see Fig.\ \ref{fig:panels}(e)).
\begin{figure*}[!t]
\includegraphics[width=\linewidth]{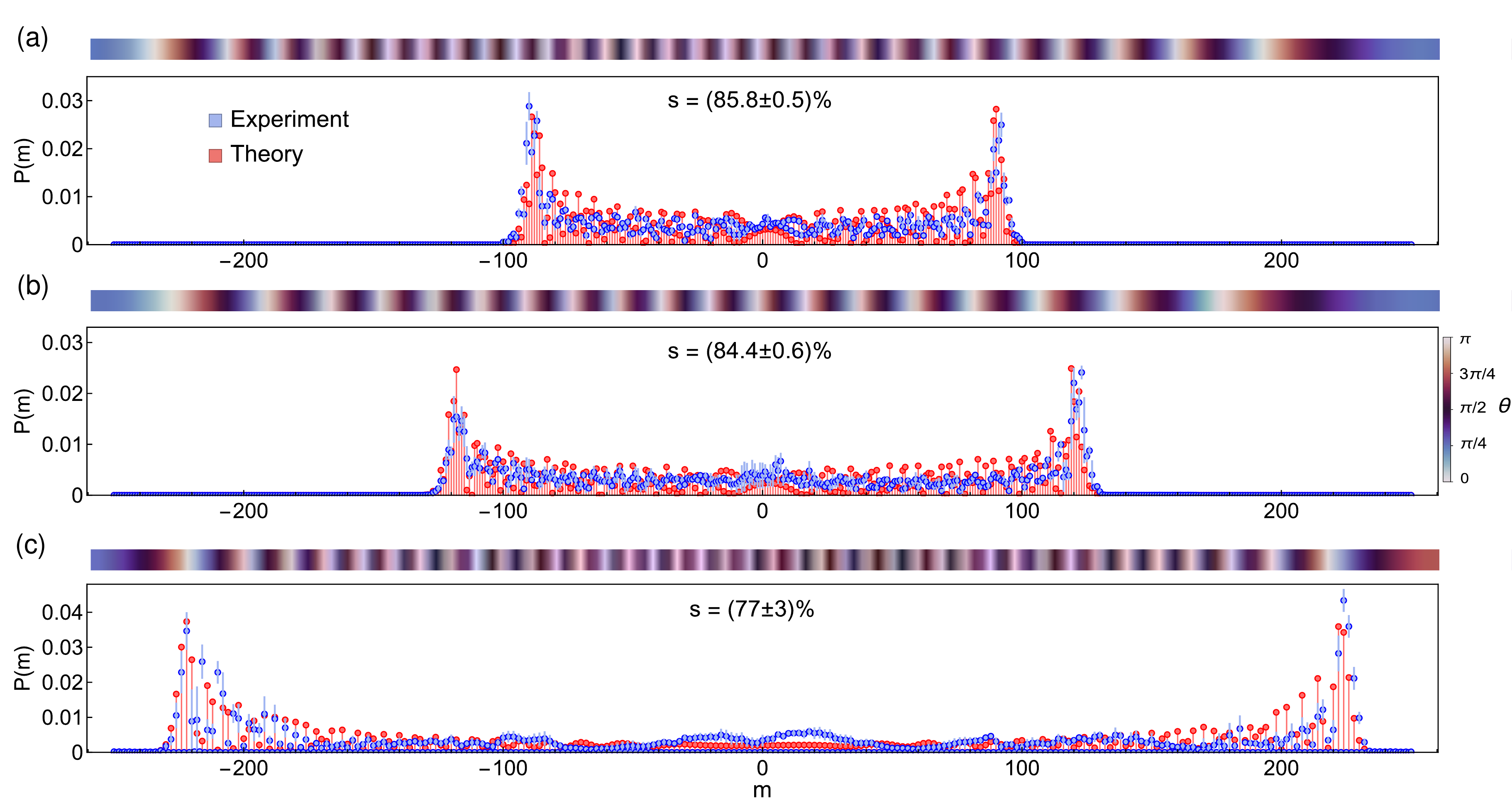}
\caption{{\bf  
Ultra-long QWs.} 
Experimental probability distributions resulting from:  
(a) $\tau=240$ steps of $U_2(7\pi/4)$; (b) $\tau=320$ steps of $U_2(7\pi/4)$; (c) $\tau=320$ steps of $U_3(\pi)$.
 Each panel 
includes the similarity $s$ and the optic-axis modulations in half a BZ ($2.5\, \text{mm}$, from $x=0$ to $x=\Lambda/2$) of the second LCMS employed in the corresponding evolution. 
}
\label{fig:many_steps}
\end{figure*}

\noindent \textbf{Ultra-long QWs.} QW simulations based on the scheme presented above are performed using the setup illustrated in Fig.\ \ref{fig:distributions}(a). A standard He-Ne laser source (with wavelength $\lambda=633$ nm) produces a linearly polarized coherent beam, which we expand via a telescopic two-lens system ($\text{L}_1$ and $\text{L}_2$) to have the final beam waist $w_0\simeq \Lambda$. A pinhole (PH) placed in the focal plane of $\text{L}_1$ acts as a spatial filter. After preparing the input coin-polarization state with a combination of a HWP and a QWP, the whole QW is realized by passing the beam across three LCMSs ($\text{Q}_1$-$\text{Q}_3$), which are set a few millimeters apart from each other.
In Fig.\ \ref{fig:distributions}(b) we report the LC patterns used to generate 20 time-steps of the single-step operator $U_1(\delta)$, whose definition is provided in the Methods. At the exit of the last metasurface, the optical field features a complex polarization pattern, with each polarization component modulated by a complex envelope with spatial period $\Lambda$. 
As a consequence, the beam is diffracted in the $xz$ plane into a superposition of the optical modes defined in Eq.\ \eqref{eqn:modes}. In the focal plane of a converging lens ($\text{L}_3$), these are focused at spatially-separated positions (see Fig.\ \ref{fig:distributions}(c)). As anticipated, 
the peak envelopes have negligible overlaps as long as $w_0\gtrsim \Lambda$ \cite{DErrico2020} (in our implementation $w_0\simeq\Lambda= 5$ mm). The optical power carried by each mode is measured using a CCD camera, and recorded images are processed to extract the walker probability distribution (see Methods). 

We start our experiments preparing the initial state $\ket{\psi_0}=\ket{m=0}\ket{L}$, that is a circularly polarized localized input. The probability distribution measured after the 20-step QW is reported as blue bars in Fig.\ \ref{fig:distributions}(d), with error bars representing statistical uncertainties obtained as discussed in the Methods. 
To quantify the agreement between our results and the expected distribution, whose values are reported as red bars in Fig.\ \ref{fig:distributions}(d), we compute the similarity ${s=(\sum_m \sqrt{P^e(m)P^i(m)})^2}$. Here, $P^e(m)$ and $P^i(m)$ represent the experimental and ideal probabilities for the walker to occupy the $m$-th site, respectively. 
For this preliminary experiment we obtain an excellent similarity, $s=93.4\pm 0.4 \%$.

We then proceed to implement much longer evolutions, generating 240 time-steps of a different QW protocol $U_2(\delta)$ (see Methods). To our knowledge, this goes far beyond previous experiments \cite{Flamini2018}, which generally reached up to 50-70 steps in absence of optical amplification mechanisms that can balance losses, as for instance in Ref.\ \cite{Regensburger2012}. 
In Fig.\ \ref{fig:many_steps}(a) we plot the measured probability distribution, which spans more than 200 lattice sites and reproduces quite accurately the predicted distribution (similarity $s=85.8\pm 0.5 \%$). In  Fig.\ \ref{fig:many_steps}(a) we also show the LC pattern of $Q_2$ (the second LCMS implementing the walk). With $\tau=240$ the LC modulations are very rapid, and actually approaching the maximum resolution presently achievable in our LCMSs (see Methods). 
To perform even longer evolutions we cascade two QW setups (each composed of three LCMSs), each implementing 160-step evolutions, thus effectively realizing 320 time-steps.
The LC optic axis patterns associated with the second 160-step LCMS are plotted in Figs.\ \ref{fig:many_steps}(b)-(c), where we also report the distributions corresponding to 320 steps of QWs $U_2(\delta)$ and $U_3(\delta)$, respectively (see Methods). 
The latter provides the most severe test for our platform, as the associated dynamics leads to a noticeably wider broadening of the walker distribution due to larger values of the underlying group velocity. The optical wavepacket spreads over more than 400 spatial modes for each polarization component, for a total of more than 800 modes. In these cases, the obtained similarities are not as high as for shorter evolutions. We expected this at least for two reasons, indeed (i) the walker wavepacket spreads over a significantly large number of sites, and (ii) unavoidable misalignments between the cells in the $x$ direction induce greater errors in the experimental evolution. 
Nonetheless, the overall experimental distributions 
remain in very good agreement with the expected results (showing similarities of at least $75\%$). In general, a more accurate reconstruction might be obtained by replacing our three-plate scheme with multiple stages exhibiting modulations with smaller spatial periods, which can be fabricated with higher accuracy. However, this would always be associated with increased optical losses.

\begin{figure}[t]
\centering
\includegraphics[width=\linewidth]{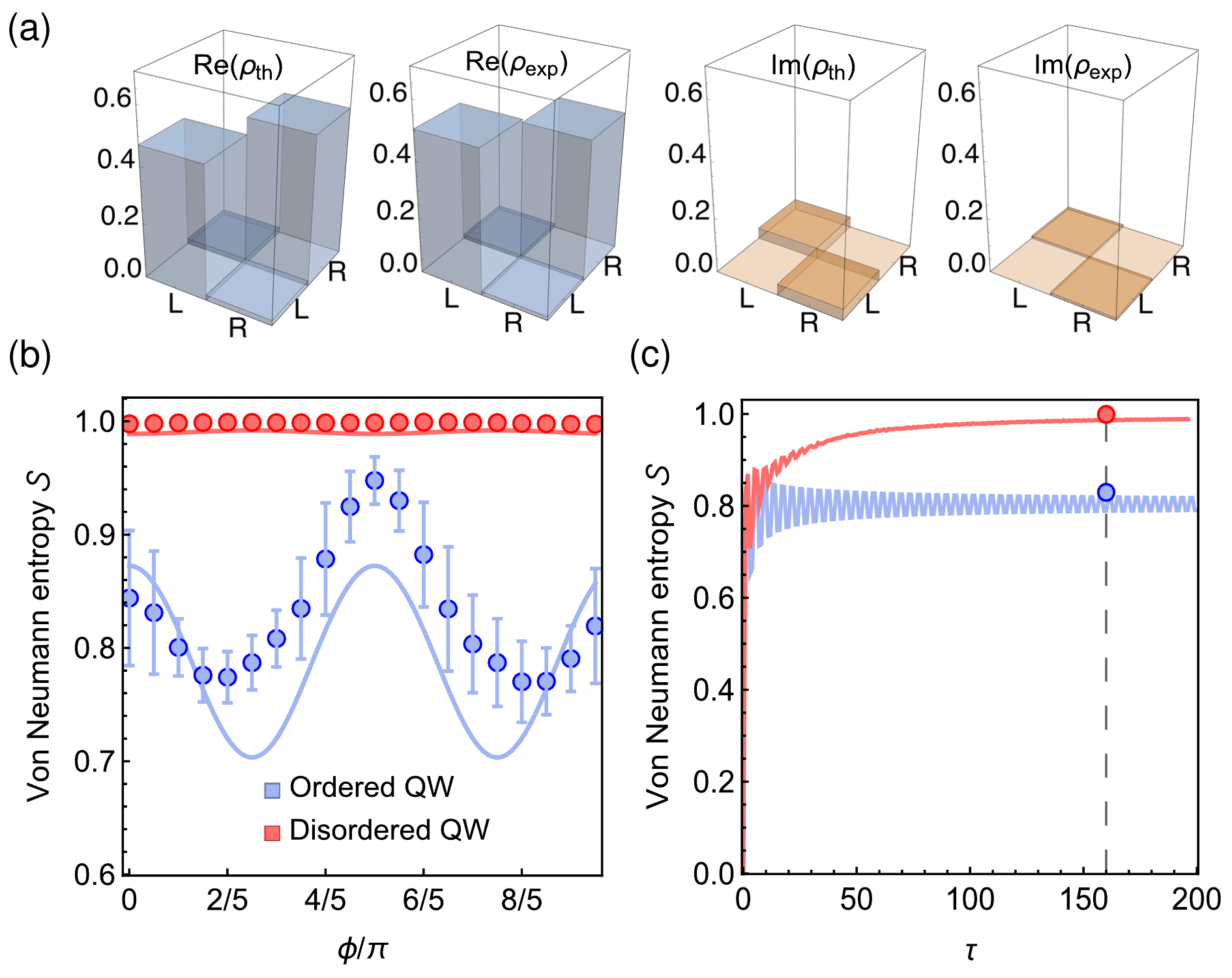}
\caption{{\bf Disordered QW as maximal entanglement generator.} 
(a) Theoretical and experimental coin density matrix for $\phi=8\pi/5$, after a disordered 160-step $U_3$ evolution.
(b) 
Von Neumann entropy $\mathcal{S}$ for QWs governed by $U_3$, for linearly polarized input states %obtained by varying the 
with azimuthal angle $\phi$ on the Poincar\'e sphere. 
Blue and red datapoints represent experimental measurements after 160 steps in the ordered and disordered case, respectively, and the curves show the expected results.
(c) The mean values of these datasets are plotted against numerical simulations of $\mathcal S(\tau)$, averaged over 100 different inputs states and over 10 realizations for each input (in the disordered configuration). Where not visible, error bars are smaller than the markers.}
\label{fig:entanglement}
\end{figure}

As shown above, our platform unlocks the realization of ultra-long (compared with the state of the art) QW evolutions in large-scale lattices, which hitherto appeared prohibitive due to the amount of required resources and optical losses.  
To provide a quantitative estimate of the efficiency of our new platform, engineering 240 steps in our setup requires three plates only. 
With a transmittance $\simeq0.85$ across each plate, the total transmittance of our setup is $(0.85)^3\simeq 0.6$, while the scheme adopted in Ref.\ \cite{DErrico2020} (which requires two plates per time-step) would transmit $(0.85)^{2*240}\simeq 10^{-34}$  of the input light.  
While other setups may have a higher single-step efficiency, 
losses will scale exponentially in every platform conceptually similar to the schemes reported in Fig.\ \ref{fig:panels}(b) or in integrated photonic systems \cite{Wang2020}.  Besides, the data we present in Fig.\ \ref{fig:many_steps} indicate that our setup guarantees a high level of coherence between the different optical modes, whose protection in long evolutions over large lattices is extremely challenging.

\noindent \textbf{Maximal entanglement generation.} Ultra-long QWs across large-scale lattices have important fundamental and technological applications. For instance, in a recent experiment long-time QWs in a 9-vertex graph enabled the measurement of the centrality ranking of directed graphs \cite{Wu2020}. However, that implementation requires a number of optical elements which grows rapidly with the system size, thereby preventing the application of such approach to large-scale systems. This is not the case in our platform, where the ability to connect an increasing number of modes within a few spin-orbit optical elements is enabled by the increasing complexity of their patterns.
As a proof of principle, we use our setup to generate maximal entanglement between the natural partitions in our system,  constituted by the coin and the walker degrees of freedom, respectively. Entanglement can be quantified in terms of the Von Neumann entropy $\mathcal{S}=-\Tr(\rho_c \log_2 \rho_c)$ of the reduced density matrix $\rho_c$ of the coin. This corresponds to the partial trace of the total density matrix (associated with the whole quantum state of the system) with respect to the spatial degree of freedom (see Methods) \cite{nielsen_chuang_2010}. In a standard 1D QW, it is known that in the long-time limit $\mathcal{S}$ converges to a finite value that is below the maximum $\mathcal S=1$. However, theoretical studies indicate that maximal entanglement generation can be achieved by adding dynamical disorder to the system and letting the input state evolve for a significantly long walk \cite{Vieira2013}. 
A first remarkable experiment was performed in a 20-step QW \cite{Wang2018}, where an optimal sequence of step operators yielding maximal entanglement was demonstrated. 
At the same time, the authors discussed that their achievement is not general, since maximal entanglement can be obtained for \textit{every} disordered sequence of step operators only in the limit of ultra-long walks. In our platform we have access to this regime, thus we can validate experimentally the theory put forward in Ref.\ \cite{Vieira2013}. 
To introduce dynamical disorder, as prescribed in the original work by Vieira {\it et al.}~\cite{Vieira2013}, we design three LCMSs which engineer the walk $\prod_{i=1}^\tau U_3(\delta_i)$, with $\delta_i$ varying randomly at each time-step in the range $[\pi-\pi/5,\pi+\pi/5]$. The result of the QW is directly
determined in terms of projective measurements of the output field on different polarization states (see Methods) \cite{Jame01_PRA}. 
We focus on linearly polarized inputs, as in our protocol the increase in entanglement is maximal for these states.
These can be expressed as $\ket{\phi}=\frac{1}{\sqrt{2}}(\ket{L}+e^{i\phi}\ket{R})$, where $\phi/2$ 
gives the angle of the polarization plane. In Fig.\ \ref{fig:entanglement}(a) we show the reconstructed density matrix for an input state with $\phi=8\pi/5$ for a single disordered realization ($\tau=160$), finding a remarkable agreement with numerical simulations. 
In Fig.\ \ref{fig:entanglement}(b) we report the measured values of the Von Neumann entropy $\mathcal{S}$ that we obtain for different linearly polarized input states, demonstrating that a disordered and prolonged evolution of {\it every} input state generates maximal entanglement $(\mathcal{S}>0.98)$. The deviations from expected values observed in the ordered case could possibly be ascribed to some form of undesirable disorder which cannot be completely extinguished in our devices, mainly related to fabrication defects. In Fig.\ \ref{fig:entanglement}(c) we observe that average values of $\mathcal S$ at $\tau=160$ are nicely positioned on the theory curves showing the dynamical evolution of the Von Neumann entropy as a function of $\tau$.

\section*{Discussion}
We presented a photonic platform capable of generating large-scale walks, corresponding at the same time to ultra-long dynamics in a Hilbert space spanned by hundreds of optical modes carrying quantized transverse momentum. 
Engineering the whole dynamics with an exceedingly limited number of optical elements
eliminates  
the need of aligning bulky optical setups, and dramatically reduces optical losses and decoherence effects. In this work, we used a classical laser source, and all these results may therefore be understood in terms of single-particle physics.
The next logical step is to study genuine multi-particle dynamics, where the photon detection for each spatial mode can be achieved by using a fiber array placed in the focal plane of the output lens, with fibers connected to standard single-photon detectors. This has been recently illustrated by some of us in a proof-of-principle demonstration, reporting QWs performed by two indistinguishable photons \cite{Esposito2022}, obtained with a standard cascaded setup \cite{DErrico2020} which permitted only 3 discrete steps in a two-dimensional walk.
On the other hand, the new platform is ideally suited for generating large-scale multi-photon QWs mixing hundreds of optical modes with long-range connectivity, and therefore represents a unique resource for cutting-edge quantum computing experiments \cite{Zhong2020,Madsen2022}.
Another natural prospect of our work is to compress two-dimensional quantum walks in a few metasurfaces. 
Although conceptually straightforward,  
complex singularities in the LC patterns  
will require a careful fabrication and an accurate alignment of the final setup. 
In parallel, we plan to investigate %the possibility to simulate  a larger class of 
QWs that do not display translation invariance, 
which may be obtained by adding symmetry-breaking layers.
While here we focused on ultra-long QWs with well-known individual-step operators, 
in the future we 
plan to directly engineer much more complex evolutions, starting from the design of the energy bands and of the associated eigenstates, realizing for instance topological flat bands in one and two spatial dimensions.
Our LCMSs are based on a technology which is conceptually
similar to the one present in commercial LC displays. The very fast industrial progresses experienced by these components could therefore straightforwardly be adapted to our setup to make it completely reconfigurable in real time in the close future. 

\section*{Methods}
%%%%%%%%%%%%%%%%%%%%%%%%%%%%%%%%
\label{app:methods}
\noindent \\ \textbf{Quantum walk protocols.} At each time-step, under the action of an operator $T$ the walker can be displaced to its neighbour sites, moving either to the right or to the left depending on the state of a two-level internal degree of freedom (the ``coin"). Between two consecutive translations, the coin state is modified through a coin rotation $W$.  
Assuming left and right circularly polarized light as the coin basis states, the coin rotation can be expressed as follows: 
\begin{equation}
W=\dfrac{1}{\sqrt{2}}
\begin{pmatrix}
1 && i\\
i && 1
\end{pmatrix},
\end{equation}
and the coin-dependent translation operator reads
\begin{equation}
T(\delta)=
\begin{pmatrix}
\cos{(\delta/2)} && i\sin{(\delta/2)}\hat{t}\\
i\sin{(\delta/2)}\hat{t}^{\dagger} && \cos{(\delta/2)}
\end{pmatrix},
\end{equation}
where the operators $\hat{t}$ and $\hat{t}^{\dagger}$ represent the forward and backward elementary translations on the lattice, acting as: $\hat{t}\ket{m}=\ket{m+1}$ and $\hat{t}^{\dagger}\ket{m}=\ket{m-1}$. The parameter $\delta$ tunes the hopping amplitudes between neighbour sites. The chosen combination of these two operators identifies a specific QW protocol $U$, also representing a single step of the process. For our 20-step QW experiment, we chose protocol $U_1(\delta)=T(\delta)W$, at $\delta=\pi$. For the 240-step and 320-step evolutions, we chose $U_2(\delta)=\sqrt{T(\delta)}W\sqrt{T(\delta)}$, at $\delta=7\pi/4$. For the longest dynamics, we also chose $U_3(\delta)=R^{\dagger}T(\delta)WR$, at $\delta=\pi$, where the additional rotation is defined as follows: 
\begin{equation}
R=
\begin{pmatrix}
\cos(\pi/8) && -i\sin(\pi/8)\\
-i\sin(\pi/8) && \cos(\pi/8)\\
\end{pmatrix}.
\end{equation}
The last protocol has been chosen also for investigating disordered evolutions. Let us note that the $U_i$ are single-step operators of a given QW. As discussed in the main text, we implement a complete $\tau$-step QW by means of only three metasurfaces, as $\mathcal{U}^\tau = (\mathcal{U}_i)^\tau =Q_{\pi/2}(\theta_3)Q_{\pi}(\theta_2)Q_{\pi/2}{(\theta_1)}$.
\\ \\ \textbf{Experimental setup.} A He-Ne laser beam ($\lambda=633$ nm) passes through a telescope system, consisting of two aspheric lenses $\text{L}_1$ and $\text{L}_2$ (with focal lengths $f_1=5$ cm and $f_2=30$ cm) and a 25$\mu$m pinhole (PH). The latter operates as a spatial filter. 
As discussed in the text, we found a convenient choice setting the beam waist $w_0\simeq\,$5 mm. A combination of a half-wave plate (HWP) and a quarter-wave plate (QWP) sets the desired input polarization. After preparing the input polarization, the beam passes through the three LCMSs implementing the full dynamics. These are located in a handy plastic box provided with external screws, allowing us to adjust their transverse displacement. This represents the crucial stage for the required precise alignment. At the exit of the box we put a lens $\text{L}_3$ (with focal length $f_3=50$ cm), Fourier-transforming light momenta into positions on the CCD camera placed in the focal plane. The camera is mounted on a motorized translation stage, as for long evolutions the full light distribution cannot be captured within a single photo shoot. 
To resolve the walker distribution, subsequent pictures are therefore merged and analyzed
(see next section). 
\\ \\ \textbf{Measuring probability distributions.} 
After experiencing the QW evolution, in the focal plane of the final lens light distributes over several spots, each one corresponding with a walker site. 
The probability $P(m)$ that the walker is found in the $m$-th site is obtained by integrating the light intensity $I(m)$ within the $m$-th light spot, and normalizing it to the total intensity, that is $P(m)=I(m)/\sum_{m'} {I(m')}$. Light intensities are integrated over a 5x5 pixels domain:
\begin{equation*}
I(m)=\sum_{i=m-2}^{m+2}\sum_{j=m-2}^{m+2}I(x_i,y_j),
\end{equation*}
$x_m$ and $y_m$ being the pixel-coordinates of the $m$-th site.
To determine the coordinates of the light spot associated with the $m$-th site, first the $m=0$ position has to be determined. This is accomplished by setting the optical retardations of the LC waveplates equal to $2\pi$. In this way, all LCMSs are ``turned off" and only the spot associated with $m=0$ is observed. Then we set the appropriate voltage for each metasurface to generate the desired evolution. Consequently, several spots appear on the camera. We computed the distances between consecutive spots and found that the average distance is $\simeq 12$ pixels for all realizations. \\ \\ \textbf{Error bars in probability distributions.} Values in each probability distribution are obtained by repeating the associated experiment four times, each measurement being performed after realigning the three LCMSs and eventually averaging the resulting intensity distributions. Error bars are estimated as the mean standard errors (MSE). This procedure allows us to take into account experimental uncertainties associated with relative misalignments between the three plates along the $x$ direction.
\\ \\ \textbf{Maximum achievable resolution and paraxial approximation.}
Since a single step of the QW process couples neighbour sites only, after $\tau$ time-steps the evolution operator contains long-range couplings, with the maximum coupling length being exactly $\tau$. In our photonic implementation, this implies that the largest transverse momentum kick is $\tau\Delta k_\perp$. Accordingly, the fastest modulation that can be present in our plates has a spatial period $\Lambda/\tau$. For a $240$-step QW evolution, the period associated with the fastest modulation is $\simeq 20 \mu$m. As the ultimate length scale at which LC molecules can be modulated is of the order of the cell thickness, this indicates that plates employed for the 240-step evolution are already approaching the maximum achievable resolution. Nevertheless, as mentioned in the text, one can stack multiple three-plate stages to implement longer evolutions. We note here that, to remain within the paraxial approximation, the transverse momentum of each mode must be much smaller than the longitudinal wavevector, i.e. $\tau \Delta k_\perp \ll 2\pi/\lambda $, where $\lambda$ is the photon wavelength. In our setup this can be approximately fulfilled up to the extreme case $\tau \simeq 800$, where $\tau\lambda/\Lambda\simeq 10^{-1}$. This number can be potentially increased by increasing $\Lambda$ and/or using photons with shorter wavelength.
\\ \\ \textbf{Computing optic-axis patterns.}
We observed that QW dynamics can be suitably mapped into position-dependent polarization transformations. These can be reproduced by means of LCMSs with a non-uniform optic-axis orientation. Here follows the description of the procedure we adopt to compute the correct values of the angles $\theta_i$ ($i=\lbrace{1,2,3\rbrace}$) to reproduce a target dynamics. The global evolution operator $\mathcal{U}^{\tau}(x)$ can be decomposed as $\mathcal{U}^{\tau}(x)=\sum_{\alpha} c_{\alpha}(x) \sigma_{\alpha}$, where $\sigma_0$ is the identity matrix and $\bm{\sigma}=\lbrace{\sigma_1,\sigma_2,\sigma_3\rbrace}$ is the vector of the three Pauli matrices. The sequence of three LCMSs $\mathcal{L}(x)$ can be analogously decomposed as $\mathcal{L}(x)=\sum_{\alpha} \ell_{\alpha}(x) \sigma_{\alpha}$. 
In order to determine the three optic-axis modulations, we have to solve the equations
\begin{equation}
\ell_{\alpha}(x)=c_{\alpha}(x), \,\,\,\,\,\,\alpha=\lbrace{0,1,2,3\rbrace}.
\label{eqn:opticaxis}
\end{equation}
\begin{figure}[!t]
\centering
\includegraphics[width=0.9\linewidth]{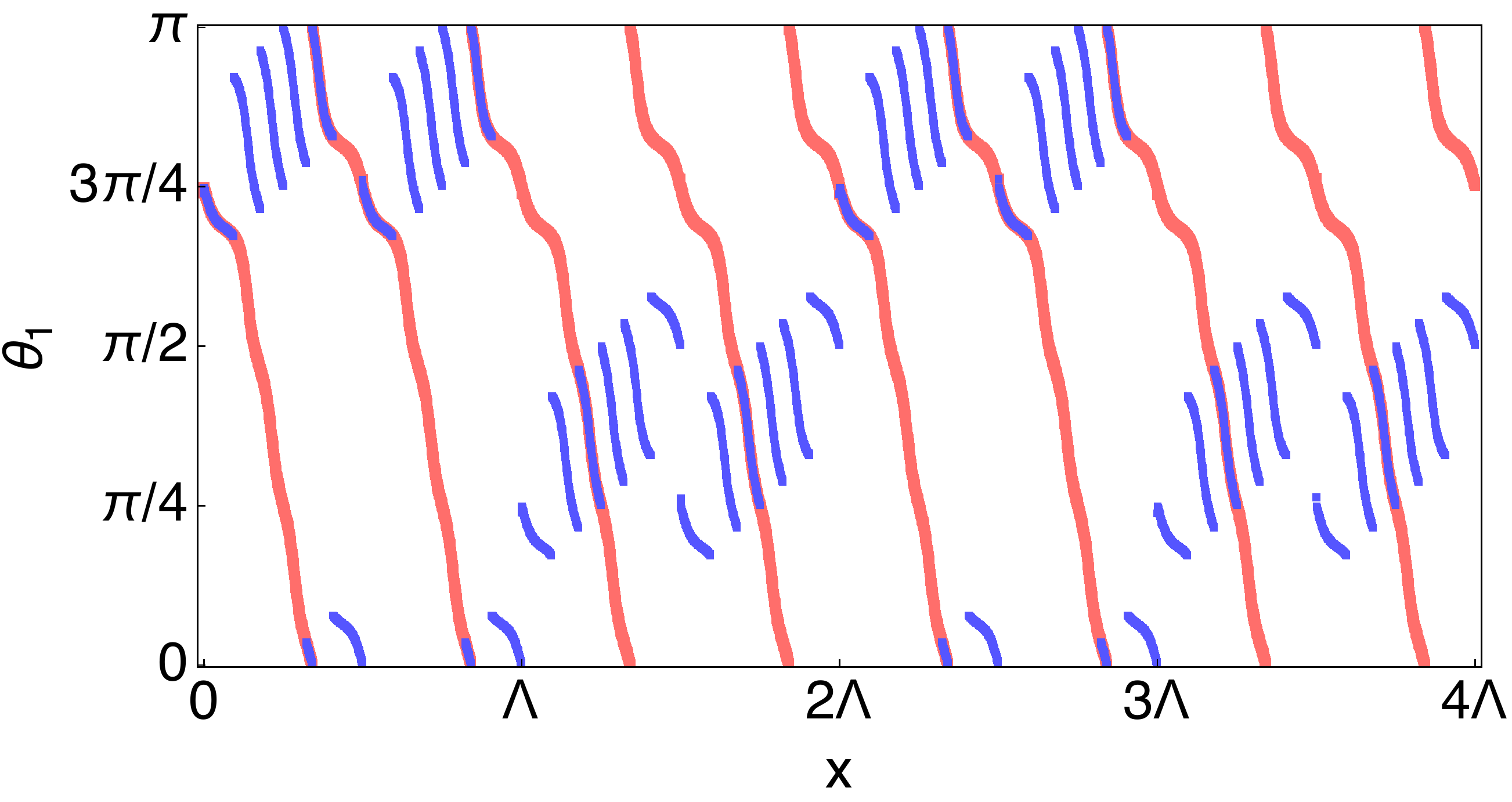}
\caption{{\bf Searching for continuous optical modulations.}
Optic-axis modulation of the first LCMS implementing $U_1^{10}(\pi)$. The blue modulation corresponds to the case when only a specific set of solutions is picked. A number of discontinuities would visibly arise in the pattern. 
Whenever a discontinuity is detected, the algorithm searches among the remaining sets of solutions until a continuous modulation (red curve) is found for the three LCMSs. 
}
\label{fig:methods}
\end{figure}
More explicitly,
\begin{equation}
\begin{cases}
-\cos(\theta_1-\theta_3)\cos(\theta_1-2\theta_2+\theta_3)=c_0\\
i \sin(\theta_1+\theta_3)\sin(\theta_1-2\theta_2+\theta_3)=c_1\\
-i \cos(\theta_1+\theta_3)\sin(\theta_1-2\theta_2+\theta_3)=c_2\\
-i \sin(\theta_1-\theta_3)\cos(\theta_1-2\theta_2+\theta_3)=c_3
\end{cases},
\label{eqn:explicitopticaxis}
\end{equation}\\
where we have omitted the dependence on $x$. 
This equation system is overcomplete, as a consequence of the unitarity of the whole process. Considering a given position $x$, eight sets of analytical solutions $\theta_i(x)$ ($i=\lbrace{1,2,3\rbrace}$) exist for the Eqs.\ \eqref{eqn:explicitopticaxis}. To compute the correct patterns to be used in the fabrication process, we fixed the length of the synthetic BZ at $\Lambda=5$ mm, discretizing it in steps of 4$\mu$m. If calculating the angles picking only one of the solutions mentioned above, we observe that computed values are not varying continuously (blue modulation in Fig.\ \ref{fig:methods}). To overcome this issue, we devise a dedicated algorithm that automatically switches the solution to be used whenever the value of the angle that has been computed features a sudden jump. This approach actually avoids jumps and yields patterns that are continuously modulated (red modulation in Fig.\ \ref{fig:methods}).
\\ \\ \textbf{Von Neumann entropy.} At any time $\tau$, the state can be written as $\vert \psi\left( \tau \right)\rangle=\sum_{m} (\alpha_m \vert L \rangle + \beta_m \vert R \rangle)\vert m \rangle$, with the normalization constraint $\sum_{m}\left(\abs{\alpha_m}^2+\abs{\beta_m}^2 \right)=1$. Accordingly, the density matrix of this pure state reads
\begin{equation}
\begin{split}
\rho =& \,\vert \psi(\tau)\rangle \langle \psi(\tau) \vert\\ 
=&\sum_{m,n} (\alpha_m \vert L \rangle + \beta_m \vert R \rangle) (\alpha_n^* \langle L \vert + \beta_n^* \langle R \vert) \vert m \rangle \langle n \vert .
\end{split}
\end{equation}
Tracing over the walker part of the state, one obtains the coin reduced density matrix
\begin{equation}
\rho_c = \, \Tr_m(\rho)=
\begin{pmatrix}
\sum_m \vert \alpha_m\vert^2 & \sum_m \alpha_m \beta_m^*\\
\sum_m \alpha_m^* \beta_m & \sum_m \vert \beta_m\vert^2
\end{pmatrix},
\end{equation}
\\from which the Von Neumann entropy ${\mathcal{S}=-\Tr(\rho_c\log_2 \rho_c)}$ can be computed.
\\ \\ \textbf{Experimental reconstruction of the reduced density matrix.} Our setup offers a natural way to trace out the walker degree of freedom, since measuring the coin reduced density matrix is equivalent to retrieving the polarization state of the whole beam. This is accomplished by measuring the beam reduced Stokes parameters $\lbrace{s_1,s_2,s_3\rbrace}$. These are obtained by recording the light intensity associated with a suitable combination of six projective measurements, performed with a quarter-wave plate, a linear polarizer and a power meter. The coin reduced density matrix can be finally reconstructed as $\rho_c=\left(\sigma_0+\sum_i s_i\sigma_i\right)/2$ \cite{Jame01_PRA}.

\section*{Statements}
%%%%%%%%%%%%%%%%%%%%%%%%%%%%%%%%%
\noindent \textbf{Acknowledgments} \\FDC, AB, LM and FC acknowledge support from the European Union Horizon 2020 program, under European Research Council (ERC) grant no. 694683 (PHOSPhOR). AD acknowledges support from: ERC AdG NOQIA; Agencia Estatal de Investigación (R\&D project CEX2019-000910-S, funded by MCIN/ AEI/10.13039/501100011033, Plan National FIDEUA PID2019-106901GB-I00, FPI, QUANTERA MAQS PCI2019-111828-2, Proyectos de I+D+I “Retos Colaboración” QUSPIN RTC2019-007196-7);  Fundacio Cellex; Fundacio Mir-Puig; Generalitat de Catalunya through the European Social Fund FEDER and CERCA program (AGAUR Grant No. 2017 SGR 134, QuantumCAT  U16-011424, co-funded by ERDF Operational Program of Catalonia 2014-2020); EU Horizon 2020 FET-OPEN OPTOlogic (Grant No 899794); National Science Centre, Poland (Symfonia Grant No. 2016/20/W/ST4/00314); European Union’s Horizon 2020 research and innovation programme under the Marie-Sklodowska-Curie grant agreement No 101029393 (STREDCH) and No 847648  (“La Caixa” Junior Leaders fellowships ID100010434: LCF/BQ/PI19/11690013, LCF/BQ/PI20/11760031, LCF/BQ/PR21/11840013).  AD further acknowledges the financial support from a fellowship granted by la Caixa Foundation (ID 100010434, fellowship code LCF/BQ/PR20/11770012).
PM was supported by grant PID2020-113565GB-C21 funded by MCIN/AEI/10.13039/501100011033, by EU FEDER Quantumcat, and by the {\it ICREA Academia} program.
\\ \\ \\ \textbf{Disclosures} \\The authors declare no conflicts of interest.
\\ \\ \\ \textbf{Data availability} \\Data underlying the results presented in this paper are not publicly available at this time but may be obtained from the authors upon reasonable request.


\begin{thebibliography}{44}%
\makeatletter
\providecommand \@ifxundefined [1]{%
 \@ifx{#1\undefined}
}%
\providecommand \@ifnum [1]{%
 \ifnum #1\expandafter \@firstoftwo
 \else \expandafter \@secondoftwo
 \fi
}%
\providecommand \@ifx [1]{%
 \ifx #1\expandafter \@firstoftwo
 \else \expandafter \@secondoftwo
 \fi
}%
\providecommand \natexlab [1]{#1}%
\providecommand \enquote  [1]{``#1''}%
\providecommand \bibnamefont  [1]{#1}%
\providecommand \bibfnamefont [1]{#1}%
\providecommand \citenamefont [1]{#1}%
\providecommand \href@noop [0]{\@secondoftwo}%
\providecommand \href [0]{\begingroup \@sanitize@url \@href}%
\providecommand \@href[1]{\@@startlink{#1}\@@href}%
\providecommand \@@href[1]{\endgroup#1\@@endlink}%
\providecommand \@sanitize@url [0]{\catcode `\\12\catcode `\$12\catcode
  `\&12\catcode `\#12\catcode `\^12\catcode `\_12\catcode `\%12\relax}%
\providecommand \@@startlink[1]{}%
\providecommand \@@endlink[0]{}%
\providecommand \url  [0]{\begingroup\@sanitize@url \@url }%
\providecommand \@url [1]{\endgroup\@href {#1}{\urlprefix }}%
\providecommand \urlprefix  [0]{URL }%
\providecommand \Eprint [0]{\href }%
\providecommand \doibase [0]{http://dx.doi.org/}%
\providecommand \selectlanguage [0]{\@gobble}%
\providecommand \bibinfo  [0]{\@secondoftwo}%
\providecommand \bibfield  [0]{\@secondoftwo}%
\providecommand \translation [1]{[#1]}%
\providecommand \BibitemOpen [0]{}%
\providecommand \bibitemStop [0]{}%
\providecommand \bibitemNoStop [0]{.\EOS\space}%
\providecommand \EOS [0]{\spacefactor3000\relax}%
\providecommand \BibitemShut  [1]{\csname bibitem#1\endcsname}%
\let\auto@bib@innerbib\@empty
%</preamble>
\bibitem [{\citenamefont {Buluta}\ and\ \citenamefont
  {Nori}(2009)}]{Buluta2009}%
  \BibitemOpen
  \bibfield  {author} {\bibinfo {author} {Buluta, I.}\ and\ \bibinfo {author}
  {Nori, F.},\ }\bibfield  {title} {\bibinfo {title} {\emph {Quantum
  Simulators}},\ }\href {\doibase 10.1126/science.1177838} {\bibfield
  {journal} {\bibinfo  {journal} {Science}\ }\textbf {\bibinfo {volume}
  {326}},\ \bibinfo {pages} {108} (\bibinfo {year} {2009})}\BibitemShut
  {NoStop}%
\bibitem [{\citenamefont {Gross}\ and\ \citenamefont
  {Bloch}(2017)}]{Gross2017}%
  \BibitemOpen
  \bibfield  {author} {\bibinfo {author} {Gross, C.}\ and\ \bibinfo {author}
  {Bloch, I.},\ }\bibfield  {title} {\bibinfo {title} {\emph {Quantum
  simulations with ultracold atoms in optical lattices}},\ }\href
  {https://www.science.org/doi/abs/10.1126/science.aal3837} {\bibfield
  {journal} {\bibinfo  {journal} {Science}\ }\textbf {\bibinfo {volume}
  {357}},\ \bibinfo {pages} {995} (\bibinfo {year} {2017})}\BibitemShut
  {NoStop}%
\bibitem [{\citenamefont {Blatt}\ and\ \citenamefont {Roos}(2012)}]{Blatt2012}%
  \BibitemOpen
  \bibfield  {author} {\bibinfo {author} {Blatt, R.}\ and\ \bibinfo {author}
  {Roos, C.~F.},\ }\bibfield  {title} {\bibinfo {title} {\emph {{Quantum
  simulations with trapped ions}}},\ }\href {\doibase 10.1038/nphys2252}
  {\bibfield  {journal} {\bibinfo  {journal} {Nat. Phys.}\ }\textbf {\bibinfo
  {volume} {8}},\ \bibinfo {pages} {277} (\bibinfo {year} {2012})}\BibitemShut
  {NoStop}%
\bibitem [{\citenamefont {Kjaergaard}\ \emph {et~al.}(2020)\citenamefont
  {Kjaergaard}, \citenamefont {Schwartz}, \citenamefont {Braum{\"{u}}ller},
  \citenamefont {Krantz}, \citenamefont {Wang}, \citenamefont {Gustavsson},\
  and\ \citenamefont {Oliver}}]{Kjaergaard2020}%
  \BibitemOpen
  \bibfield  {author} {\bibinfo {author} {Kjaergaard, M.}, \bibinfo {author}
  {Schwartz, M.~E.}, \bibinfo {author} {Braum{\"{u}}ller, J.}, \bibinfo
  {author} {Krantz, P.}, \bibinfo {author} {Wang, J.~I.}, \bibinfo {author}
  {Gustavsson, S.}\ and\ \bibinfo {author} {Oliver, W.~D.},\ }\bibfield
  {title} {\bibinfo {title} {\emph {{Superconducting Qubits: Current State of
  Play}}},\ }\href {\doibase 10.1146/annurev-conmatphys-031119-050605}
  {\bibfield  {journal} {\bibinfo  {journal} {Annu. Rev. Condens. Matter
  Phys.}\ }\textbf {\bibinfo {volume} {11}},\ \bibinfo {pages} {369} (\bibinfo
  {year} {2020})}\BibitemShut {NoStop}%
\bibitem [{\citenamefont {Wang}\ \emph {et~al.}(2020)\citenamefont {Wang},
  \citenamefont {Sciarrino}, \citenamefont {Laing},\ and\ \citenamefont
  {Thompson}}]{Wang2020}%
  \BibitemOpen
  \bibfield  {author} {\bibinfo {author} {Wang, J.}, \bibinfo {author}
  {Sciarrino, F.}, \bibinfo {author} {Laing, A.}\ and\ \bibinfo {author}
  {Thompson, M.~G.},\ }\bibfield  {title} {\bibinfo {title} {\emph {{Integrated
  photonic quantum technologies}}},\ }\href {\doibase
  http://dx.doi.org/10.1038/s41566-019-0532-1} {\bibfield  {journal} {\bibinfo
  {journal} {Nat. Photonics}\ }\textbf {\bibinfo {volume} {14}},\ \bibinfo
  {pages} {273} (\bibinfo {year} {2020})}\BibitemShut {NoStop}%
\bibitem [{\citenamefont {Zhong}\ \emph {et~al.}(2020)\citenamefont {Zhong},
  \citenamefont {Wang}, \citenamefont {Deng}, \citenamefont {Chen},
  \citenamefont {Peng}, \citenamefont {Luo}, \citenamefont {Qin}, \citenamefont
  {Wu}, \citenamefont {Ding}, \citenamefont {Hu}, \citenamefont {Hu},
  \citenamefont {Yang}, \citenamefont {Zhang}, \citenamefont {Li},
  \citenamefont {Li}, \citenamefont {Jiang}, \citenamefont {Gan}, \citenamefont
  {Yang}, \citenamefont {You}, \citenamefont {Wang}, \citenamefont {Li},
  \citenamefont {Liu}, \citenamefont {Lu},\ and\ \citenamefont
  {Pan}}]{Zhong2020}%
  \BibitemOpen
  \bibfield  {author} {\bibinfo {author} {Zhong, H.-S.} \emph {et~al.},\
  }\bibfield  {title} {\bibinfo {title} {\emph {{Quantum computational
  advantage using photons}}},\ }\href {\doibase
  http://dx.doi.org/10.1126/science.abe8770} {\bibfield  {journal} {\bibinfo
  {journal} {Science}\ }\textbf {\bibinfo {volume} {370}},\ \bibinfo {pages}
  {1460} (\bibinfo {year} {2020})}\BibitemShut {NoStop}%
\bibitem [{\citenamefont {Slussarenko}\ and\ \citenamefont
  {Pryde}(2019)}]{Slussarenko2019}%
  \BibitemOpen
  \bibfield  {author} {\bibinfo {author} {Slussarenko, S.}\ and\ \bibinfo
  {author} {Pryde, G.~J.},\ }\bibfield  {title} {\bibinfo {title} {\emph
  {{Photonic quantum information processing: A concise review}}},\ }\href
  {\doibase http://dx.doi.org/10.1063/1.5115814} {\bibfield  {journal}
  {\bibinfo  {journal} {Appl. Phys. Rev.}\ }\textbf {\bibinfo {volume} {6}},\
  \bibinfo {pages} {041303} (\bibinfo {year} {2019})}\BibitemShut {NoStop}%
\bibitem [{\citenamefont {Bogaerts}\ \emph {et~al.}(2020)\citenamefont
  {Bogaerts}, \citenamefont {P{\'{e}}rez}, \citenamefont {Capmany},
  \citenamefont {Miller}, \citenamefont {Poon}, \citenamefont {Englund},
  \citenamefont {Morichetti},\ and\ \citenamefont {Melloni}}]{Bogaerts2020}%
  \BibitemOpen
  \bibfield  {author} {\bibinfo {author} {Bogaerts, W.}, \bibinfo {author}
  {P{\'{e}}rez, D.}, \bibinfo {author} {Capmany, J.}, \bibinfo {author}
  {Miller, D. A.~B.}, \bibinfo {author} {Poon, J.}, \bibinfo {author} {Englund,
  D.}, \bibinfo {author} {Morichetti, F.}\ and\ \bibinfo {author} {Melloni,
  A.},\ }\bibfield  {title} {\bibinfo {title} {\emph {{Programmable photonic
  circuits}}},\ }\href {\doibase 10.1038/s41586-020-2764-0} {\bibfield
  {journal} {\bibinfo  {journal} {Nature}\ }\textbf {\bibinfo {volume} {586}},\
  \bibinfo {pages} {207} (\bibinfo {year} {2020})}\BibitemShut {NoStop}%
\bibitem [{\citenamefont {Blais}\ \emph {et~al.}(2020)\citenamefont {Blais},
  \citenamefont {Girvin},\ and\ \citenamefont {Oliver}}]{Blais2020}%
  \BibitemOpen
  \bibfield  {author} {\bibinfo {author} {Blais, A.}, \bibinfo {author}
  {Girvin, S.~M.}\ and\ \bibinfo {author} {Oliver, W.~D.},\ }\bibfield  {title}
  {\bibinfo {title} {\emph {{Quantum information processing and quantum optics
  with circuit quantum electrodynamics}}},\ }\href {\doibase
  http://dx.doi.org/10.1038/s41567-020-0806-z} {\bibfield  {journal} {\bibinfo
  {journal} {Nat. Phys.}\ }\textbf {\bibinfo {volume} {16}},\ \bibinfo {pages}
  {247} (\bibinfo {year} {2020})}\BibitemShut {NoStop}%
\bibitem [{\citenamefont {Pezz{\`{e}}}\ \emph {et~al.}(2018)\citenamefont
  {Pezz{\`{e}}}, \citenamefont {Smerzi}, \citenamefont {Oberthaler},
  \citenamefont {Schmied},\ and\ \citenamefont {Treutlein}}]{Pezze2018}%
  \BibitemOpen
  \bibfield  {author} {\bibinfo {author} {Pezz{\`{e}}, L.}, \bibinfo {author}
  {Smerzi, A.}, \bibinfo {author} {Oberthaler, M.~K.}, \bibinfo {author}
  {Schmied, R.}\ and\ \bibinfo {author} {Treutlein, P.},\ }\bibfield  {title}
  {\bibinfo {title} {\emph {{Quantum metrology with nonclassical states of
  atomic ensembles}}},\ }\href {\doibase
  http://dx.doi.org/10.1103/RevModPhys.90.035005} {\bibfield  {journal}
  {\bibinfo  {journal} {Rev. Mod. Phys.}\ }\textbf {\bibinfo {volume} {90}},\
  \bibinfo {pages} {035005} (\bibinfo {year} {2018})}\BibitemShut {NoStop}%
\bibitem [{\citenamefont {Venegas-Andraca}(2012)}]{Venegas2012}%
  \BibitemOpen
  \bibfield  {author} {\bibinfo {author} {Venegas-Andraca, S.~E.},\ }\bibfield
  {title} {\bibinfo {title} {\emph {Quantum walks: a comprehensive review}},\
  }\href {\doibase 10.1007/s11128-012-0432-5} {\bibfield  {journal} {\bibinfo
  {journal} {Quantum Inf. Process.}\ }\textbf {\bibinfo {volume} {11}},\
  \bibinfo {pages} {1015} (\bibinfo {year} {2012})}\BibitemShut {NoStop}%
\bibitem [{\citenamefont {Gong}\ \emph {et~al.}(2021)\citenamefont {Gong},
  \citenamefont {Wang}, \citenamefont {Zha}, \citenamefont {Chen},
  \citenamefont {Huang}, \citenamefont {Wu}, \citenamefont {Zhu}, \citenamefont
  {Zhao}, \citenamefont {Li}, \citenamefont {Guo}, \citenamefont {Qian},
  \citenamefont {Ye}, \citenamefont {Chen}, \citenamefont {Ying}, \citenamefont
  {Yu}, \citenamefont {Fan}, \citenamefont {Wu}, \citenamefont {Su},
  \citenamefont {Deng}, \citenamefont {Rong}, \citenamefont {Zhang},
  \citenamefont {Cao}, \citenamefont {Lin}, \citenamefont {Xu}, \citenamefont
  {Sun}, \citenamefont {Guo}, \citenamefont {Li}, \citenamefont {Liang},
  \citenamefont {Bastidas}, \citenamefont {Nemoto}, \citenamefont {Munro},
  \citenamefont {Huo}, \citenamefont {Lu}, \citenamefont {Peng}, \citenamefont
  {Zhu},\ and\ \citenamefont {Pan}}]{Gong2021}%
  \BibitemOpen
  \bibfield  {author} {\bibinfo {author} {Gong, M.} \emph {et~al.},\ }\bibfield
   {title} {\bibinfo {title} {\emph {Quantum walks on a programmable
  two-dimensional 62-qubit superconducting processor}},\ }\href {\doibase
  http://dx.doi.org/10.1126/science.abg7812} {\bibfield  {journal} {\bibinfo
  {journal} {Science}\ }\textbf {\bibinfo {volume} {372}},\ \bibinfo {pages}
  {948} (\bibinfo {year} {2021})}\BibitemShut {NoStop}%
\bibitem [{\citenamefont {Mare{\v{s}}}\ \emph {et~al.}(2020)\citenamefont
  {Mare{\v{s}}}, \citenamefont {Novotn{\'{y}}},\ and\ \citenamefont
  {Jex}}]{Mares2020}%
  \BibitemOpen
  \bibfield  {author} {\bibinfo {author} {Mare{\v{s}}, J.}, \bibinfo {author}
  {Novotn{\'{y}}, J.}\ and\ \bibinfo {author} {Jex, I.},\ }\bibfield  {title}
  {\bibinfo {title} {\emph {Quantum walk transport on carbon nanotube
  structures}},\ }\href {https://doi.org/10.1016/j.physleta.2020.126302}
  {\bibfield  {journal} {\bibinfo  {journal} {Phys. Lett. A}\ }\textbf
  {\bibinfo {volume} {384}},\ \bibinfo {pages} {126302} (\bibinfo {year}
  {2020})}\BibitemShut {NoStop}%
\bibitem [{\citenamefont {Kitagawa}\ \emph {et~al.}(2010)\citenamefont
  {Kitagawa}, \citenamefont {Rudner}, \citenamefont {Berg},\ and\ \citenamefont
  {Demler}}]{Kitagawa2010}%
  \BibitemOpen
  \bibfield  {author} {\bibinfo {author} {Kitagawa, T.}, \bibinfo {author}
  {Rudner, M.~S.}, \bibinfo {author} {Berg, E.}\ and\ \bibinfo {author}
  {Demler, E.},\ }\bibfield  {title} {\bibinfo {title} {\emph {{Exploring
  Topological Phases With Quantum Walks}}},\ }\href {\doibase
  http://dx.doi.org/10.1103/PhysRevA.82.033429} {\bibfield  {journal} {\bibinfo
   {journal} {Phys. Rev. A}\ }\textbf {\bibinfo {volume} {82}},\ \bibinfo
  {pages} {033429} (\bibinfo {year} {2010})}\BibitemShut {NoStop}%
\bibitem [{\citenamefont {Schreiber}\ \emph {et~al.}(2012)\citenamefont
  {Schreiber}, \citenamefont {G{\'{a}}bris}, \citenamefont {Rohde},
  \citenamefont {Laiho}, \citenamefont {{\v{S}}tefa{\v n}{\'{a}}k},
  \citenamefont {Poto{\v{c}}ek}, \citenamefont {Hamilton}, \citenamefont
  {Jex},\ and\ \citenamefont {Silberhorn}}]{Schreiber2012}%
  \BibitemOpen
  \bibfield  {author} {\bibinfo {author} {Schreiber, A.}, \bibinfo {author}
  {G{\'{a}}bris, A.}, \bibinfo {author} {Rohde, P.~P.}, \bibinfo {author}
  {Laiho, K.}, \bibinfo {author} {{\v{S}}tefa{\v n}{\'{a}}k, M.}, \bibinfo
  {author} {Poto{\v{c}}ek, V.}, \bibinfo {author} {Hamilton, C.}, \bibinfo
  {author} {Jex, I.}\ and\ \bibinfo {author} {Silberhorn, C.},\ }\bibfield
  {title} {\bibinfo {title} {\emph {{A 2D Quantum Walk Simulation of
  Two-Particle Dynamics}}},\ }\href {\doibase 10.1126/science.1218448}
  {\bibfield  {journal} {\bibinfo  {journal} {Science}\ }\textbf {\bibinfo
  {volume} {336}},\ \bibinfo {pages} {55} (\bibinfo {year} {2012})}\BibitemShut
  {NoStop}%
\bibitem [{\citenamefont {Broome}\ \emph {et~al.}(2010)\citenamefont {Broome},
  \citenamefont {Fedrizzi}, \citenamefont {Lanyon}, \citenamefont {Kassal},
  \citenamefont {Aspuru-Guzik},\ and\ \citenamefont {White}}]{Broo10_PRL}%
  \BibitemOpen
  \bibfield  {author} {\bibinfo {author} {Broome, M.~A.}, \bibinfo {author}
  {Fedrizzi, A.}, \bibinfo {author} {Lanyon, B.~P.}, \bibinfo {author} {Kassal,
  I.}, \bibinfo {author} {Aspuru-Guzik, A.}\ and\ \bibinfo {author} {White,
  A.~G.},\ }\bibfield  {title} {\bibinfo {title} {\emph {Discrete Single-Photon
  Quantum Walks with Tunable Decoherence}},\ }\href {\doibase
  http://dx.doi.org/10.1103/PhysRevLett.104.153602} {\bibfield  {journal}
  {\bibinfo  {journal} {Phys. Rev. Lett.}\ }\textbf {\bibinfo {volume} {104}},\
  \bibinfo {pages} {153602} (\bibinfo {year} {2010})}\BibitemShut {NoStop}%
\bibitem [{\citenamefont {Cardano}\ \emph {et~al.}(2017)\citenamefont
  {Cardano}, \citenamefont {D'Errico}, \citenamefont {Dauphin}, \citenamefont
  {Maffei}, \citenamefont {Piccirillo}, \citenamefont {de~Lisio}, \citenamefont
  {{De Filippis}}, \citenamefont {Cataudella}, \citenamefont {Santamato},
  \citenamefont {Marrucci}, \citenamefont {Lewenstein},\ and\ \citenamefont
  {Massignan}}]{Cardano2017}%
  \BibitemOpen
  \bibfield  {author} {\bibinfo {author} {Cardano, F.} \emph {et~al.},\
  }\bibfield  {title} {\bibinfo {title} {\emph {{Detection of Zak phases and
  topological invariants in a chiral quantum walk of twisted photons}}},\
  }\href {\doibase https://doi.org/10.1038/ncomms15516} {\bibfield  {journal}
  {\bibinfo  {journal} {Nat. Commun.}\ }\textbf {\bibinfo {volume} {8}},\
  \bibinfo {pages} {15516} (\bibinfo {year} {2017})}\BibitemShut {NoStop}%
\bibitem [{\citenamefont {Defienne}\ \emph {et~al.}(2016)\citenamefont
  {Defienne}, \citenamefont {Barbieri}, \citenamefont {Walmsley}, \citenamefont
  {Smith},\ and\ \citenamefont {Gigan}}]{Defienne2016}%
  \BibitemOpen
  \bibfield  {author} {\bibinfo {author} {Defienne, H.}, \bibinfo {author}
  {Barbieri, M.}, \bibinfo {author} {Walmsley, I.~A.}, \bibinfo {author}
  {Smith, B.~J.}\ and\ \bibinfo {author} {Gigan, S.},\ }\bibfield  {title}
  {\bibinfo {title} {\emph {{Two-photon quantum walk in a multimode fiber}}},\
  }\href {\doibase 10.1126/sciadv.1501054} {\bibfield  {journal} {\bibinfo
  {journal} {Sci. Adv.}\ }\textbf {\bibinfo {volume} {2}},\ \bibinfo {pages}
  {e1501054} (\bibinfo {year} {2016})}\BibitemShut {NoStop}%
\bibitem [{\citenamefont {Tang}\ \emph {et~al.}(2018)\citenamefont {Tang},
  \citenamefont {{Di Franco}}, \citenamefont {Shi}, \citenamefont {He},
  \citenamefont {Feng}, \citenamefont {Gao}, \citenamefont {Sun}, \citenamefont
  {Li}, \citenamefont {Jiao}, \citenamefont {Wang}, \citenamefont {Kim},\ and\
  \citenamefont {Jin}}]{Tang2018b}%
  \BibitemOpen
  \bibfield  {author} {\bibinfo {author} {Tang, H.} \emph {et~al.},\ }\bibfield
   {title} {\bibinfo {title} {\emph {Experimental quantum fast hitting on
  hexagonal graphs}},\ }\href {\doibase 10.1038/s41566-018-0282-5} {\bibfield
  {journal} {\bibinfo  {journal} {Nat. Photonics}\ }\textbf {\bibinfo {volume}
  {12}},\ \bibinfo {pages} {754} (\bibinfo {year} {2018})}\BibitemShut
  {NoStop}%
\bibitem [{\citenamefont {Brod}\ \emph {et~al.}(2019)\citenamefont {Brod},
  \citenamefont {Galv{\~{a}}o}, \citenamefont {Crespi}, \citenamefont
  {Osellame},\ and\ \citenamefont {Spagnolo}}]{Brod2019}%
  \BibitemOpen
  \bibfield  {author} {\bibinfo {author} {Brod, D.~J.}, \bibinfo {author}
  {Galv{\~{a}}o, E.~F.}, \bibinfo {author} {Crespi, A.}, \bibinfo {author}
  {Osellame, R.}\ and\ \bibinfo {author} {Spagnolo, N.},\ }\bibfield  {title}
  {\bibinfo {title} {\emph {{Photonic implementation of boson sampling: a
  review}}},\ }\href {\doibase 10.1117/1.AP.1.3.034001} {\bibfield  {journal}
  {\bibinfo  {journal} {Adv. Photonics}\ }\textbf {\bibinfo {volume} {1}},\
  \bibinfo {pages} {1} (\bibinfo {year} {2019})}\BibitemShut {NoStop}%
\bibitem [{\citenamefont {Madsen}\ \emph {et~al.}(2022)\citenamefont {Madsen},
  \citenamefont {Laudenbach}, \citenamefont {Askarani}, \citenamefont
  {Rortais}, \citenamefont {Vincent}, \citenamefont {Bulmer}, \citenamefont
  {Miatto}, \citenamefont {Neuhaus}, \citenamefont {Helt}, \citenamefont
  {Collins}, \citenamefont {Lita}, \citenamefont {Gerrits}, \citenamefont
  {Nam}, \citenamefont {Vaidya}, \citenamefont {Menotti}, \citenamefont
  {Dhand}, \citenamefont {Vernon}, \citenamefont {Quesada},\ and\ \citenamefont
  {Lavoie}}]{Madsen2022}%
  \BibitemOpen
  \bibfield  {author} {\bibinfo {author} {Madsen, L.~S.} \emph {et~al.},\
  }\bibfield  {title} {\bibinfo {title} {\emph {{Quantum computational
  advantage with a programmable photonic processor}}},\ }\href {\doibase
  https://dx.doi.org/10.1038/s41586-022-04725-x} {\bibfield  {journal}
  {\bibinfo  {journal} {Nature}\ }\textbf {\bibinfo {volume} {606}},\ \bibinfo
  {pages} {75} (\bibinfo {year} {2022})}\BibitemShut {NoStop}%
\bibitem [{\citenamefont {Arrazola}\ \emph {et~al.}(2021)\citenamefont
  {Arrazola}, \citenamefont {Bergholm}, \citenamefont {Br{\'{a}}dler},
  \citenamefont {Bromley}, \citenamefont {Collins}, \citenamefont {Dhand},
  \citenamefont {Fumagalli}, \citenamefont {Gerrits}, \citenamefont {Goussev},
  \citenamefont {Helt}, \citenamefont {Hundal}, \citenamefont {Isacsson},
  \citenamefont {Israel}, \citenamefont {Izaac}, \citenamefont {Jahangiri},
  \citenamefont {Janik}, \citenamefont {Killoran}, \citenamefont {Kumar},
  \citenamefont {Lavoie}, \citenamefont {Lita}, \citenamefont {Mahler},
  \citenamefont {Menotti}, \citenamefont {Morrison}, \citenamefont {Nam},
  \citenamefont {Neuhaus}, \citenamefont {Qi}, \citenamefont {Quesada},
  \citenamefont {Repingon}, \citenamefont {Sabapathy}, \citenamefont {Schuld},
  \citenamefont {Su}, \citenamefont {Swinarton}, \citenamefont {Sz{\'{a}}va},
  \citenamefont {Tan}, \citenamefont {Tan}, \citenamefont {Vaidya},
  \citenamefont {Vernon}, \citenamefont {Zabaneh},\ and\ \citenamefont
  {Zhang}}]{Arrazola2021}%
  \BibitemOpen
  \bibfield  {author} {\bibinfo {author} {Arrazola, J.~M.} \emph {et~al.},\
  }\bibfield  {title} {\bibinfo {title} {\emph {{Quantum circuits with many
  photons on a programmable nanophotonic chip}}},\ }\href {\doibase
  http://dx.doi.org/10.1038/s41586-021-03202-1} {\bibfield  {journal} {\bibinfo
   {journal} {Nature}\ }\textbf {\bibinfo {volume} {591}},\ \bibinfo {pages}
  {54} (\bibinfo {year} {2021})}\BibitemShut {NoStop}%
\bibitem [{\citenamefont {D'Errico}\ \emph {et~al.}(2020)\citenamefont
  {D'Errico}, \citenamefont {Cardano}, \citenamefont {Maffei}, \citenamefont
  {Dauphin}, \citenamefont {Barboza}, \citenamefont {Esposito}, \citenamefont
  {Piccirillo}, \citenamefont {Lewenstein}, \citenamefont {Massignan},\ and\
  \citenamefont {Marrucci}}]{DErrico2020}%
  \BibitemOpen
  \bibfield  {author} {\bibinfo {author} {D'Errico, A.}, \bibinfo {author}
  {Cardano, F.}, \bibinfo {author} {Maffei, M.}, \bibinfo {author} {Dauphin,
  A.}, \bibinfo {author} {Barboza, R.}, \bibinfo {author} {Esposito, C.},
  \bibinfo {author} {Piccirillo, B.}, \bibinfo {author} {Lewenstein, M.},
  \bibinfo {author} {Massignan, P.}\ and\ \bibinfo {author} {Marrucci, L.},\
  }\bibfield  {title} {\bibinfo {title} {\emph {{Two-dimensional topological
  quantum walks in the momentum space of structured light}}},\ }\href {\doibase
  http://dx.doi.org/10.1364/OPTICA.365028} {\bibfield  {journal} {\bibinfo
  {journal} {Optica}\ }\textbf {\bibinfo {volume} {7}},\ \bibinfo {pages} {108}
  (\bibinfo {year} {2020})}\BibitemShut {NoStop}%
\bibitem [{\citenamefont {Plachta}\ \emph {et~al.}(2022)\citenamefont
  {Plachta}, \citenamefont {Hiekkam{\"{a}}ki}, \citenamefont {Yakaryılmaz},\
  and\ \citenamefont {Fickler}}]{Plachta2022}%
  \BibitemOpen
  \bibfield  {author} {\bibinfo {author} {Plachta, S. Z.~D.}, \bibinfo {author}
  {Hiekkam{\"{a}}ki, M.}, \bibinfo {author} {Yakaryılmaz, A.}\ and\ \bibinfo
  {author} {Fickler, R.},\ }\bibfield  {title} {\bibinfo {title} {\emph
  {{Quantum advantage using high-dimensional twisted photons as quantum finite
  automata}}},\ }\href {\doibase 10.22331/q-2022-06-30-752} {\bibfield
  {journal} {\bibinfo  {journal} {Quantum}\ }\textbf {\bibinfo {volume} {6}},\
  \bibinfo {pages} {752} (\bibinfo {year} {2022})}\BibitemShut {NoStop}%
\bibitem [{\citenamefont {Goel}\ \emph {et~al.}(2022)\citenamefont {Goel},
  \citenamefont {Leedumrongwatthanakun}, \citenamefont {Valencia},
  \citenamefont {McCutcheon}, \citenamefont {Conti}, \citenamefont {Pinkse},\
  and\ \citenamefont {Malik}}]{Goel2022}%
  \BibitemOpen
  \bibfield  {author} {\bibinfo {author} {Goel, S.}, \bibinfo {author}
  {Leedumrongwatthanakun, S.}, \bibinfo {author} {Valencia, N.~H.}, \bibinfo
  {author} {McCutcheon, W.}, \bibinfo {author} {Conti, C.}, \bibinfo {author}
  {Pinkse, P. W.~H.}\ and\ \bibinfo {author} {Malik, M.},\ }\bibfield  {title}
  {\bibinfo {title} {\emph {{Inverse-design of high-dimensional quantum optical
  circuits in a complex medium}}},\ }\href {http://arxiv.org/abs/2204.00578}
  {\bibfield  {journal} {\bibinfo  {journal} {arXiv:2204.00578}\ } (\bibinfo
  {year} {2022})}\BibitemShut {NoStop}%
\bibitem [{\citenamefont {Forbes}\ \emph {et~al.}(2021)\citenamefont {Forbes},
  \citenamefont {de~Oliveira},\ and\ \citenamefont {Dennis}}]{Forbes2021}%
  \BibitemOpen
  \bibfield  {author} {\bibinfo {author} {Forbes, A.}, \bibinfo {author}
  {de~Oliveira, M.}\ and\ \bibinfo {author} {Dennis, M.~R.},\ }\bibfield
  {title} {\bibinfo {title} {\emph {{Structured light}}},\ }\href {\doibase
  10.1038/s41566-021-00780-4} {\bibfield  {journal} {\bibinfo  {journal} {Nat.
  Photonics}\ }\textbf {\bibinfo {volume} {15}},\ \bibinfo {pages} {253}
  (\bibinfo {year} {2021})}\BibitemShut {NoStop}%
\bibitem [{\citenamefont {Cardano}\ and\ \citenamefont
  {Marrucci}(2015)}]{Cardano2015b}%
  \BibitemOpen
  \bibfield  {author} {\bibinfo {author} {Cardano, F.}\ and\ \bibinfo {author}
  {Marrucci, L.},\ }\bibfield  {title} {\bibinfo {title} {\emph {{Spin--orbit
  photonics}}},\ }\href {\doibase 10.1038/nphoton.2015.232} {\bibfield
  {journal} {\bibinfo  {journal} {Nat. Photonics}\ }\textbf {\bibinfo {volume}
  {9}},\ \bibinfo {pages} {776} (\bibinfo {year} {2015})}\BibitemShut {NoStop}%
\bibitem [{\citenamefont {Rubano}\ \emph {et~al.}(2019)\citenamefont {Rubano},
  \citenamefont {Cardano}, \citenamefont {Piccirillo},\ and\ \citenamefont
  {Marrucci}}]{Rubano2019}%
  \BibitemOpen
  \bibfield  {author} {\bibinfo {author} {Rubano, A.}, \bibinfo {author}
  {Cardano, F.}, \bibinfo {author} {Piccirillo, B.}\ and\ \bibinfo {author}
  {Marrucci, L.},\ }\bibfield  {title} {\bibinfo {title} {\emph {{Q-plate
  technology: a progress review [Invited]}}},\ }\href {\doibase
  http://dx.doi.org/10.1364/JOSAB.36.000D70} {\bibfield  {journal} {\bibinfo
  {journal} {J. Opt. Soc. Am. B}\ }\textbf {\bibinfo {volume} {36}},\ \bibinfo
  {pages} {D70} (\bibinfo {year} {2019})}\BibitemShut {NoStop}%
\bibitem [{\citenamefont {Celi}\ \emph {et~al.}(2014)\citenamefont {Celi},
  \citenamefont {Massignan}, \citenamefont {Ruseckas}, \citenamefont {Goldman},
  \citenamefont {Spielman}, \citenamefont {Juzeli{\=u}nas},\ and\ \citenamefont
  {Lewenstein}}]{Celi2014}%
  \BibitemOpen
  \bibfield  {author} {\bibinfo {author} {Celi, A.}, \bibinfo {author}
  {Massignan, P.}, \bibinfo {author} {Ruseckas, J.}, \bibinfo {author}
  {Goldman, N.}, \bibinfo {author} {Spielman, I.~B.}, \bibinfo {author}
  {Juzeli{\=u}nas, G.}\ and\ \bibinfo {author} {Lewenstein, M.},\ }\bibfield
  {title} {\bibinfo {title} {\emph {{Synthetic Gauge Fields in Synthetic
  Dimensions}}},\ }\href {\doibase 10.1103/PhysRevLett.112.043001} {\bibfield
  {journal} {\bibinfo  {journal} {Phys. Rev. Lett.}\ }\textbf {\bibinfo
  {volume} {112}},\ \bibinfo {pages} {043001} (\bibinfo {year}
  {2014})}\BibitemShut {NoStop}%
\bibitem [{\citenamefont {Yuan}\ \emph {et~al.}(2018)\citenamefont {Yuan},
  \citenamefont {Lin}, \citenamefont {Xiao},\ and\ \citenamefont
  {Fan}}]{Yuan2018}%
  \BibitemOpen
  \bibfield  {author} {\bibinfo {author} {Yuan, L.}, \bibinfo {author} {Lin,
  Q.}, \bibinfo {author} {Xiao, M.}\ and\ \bibinfo {author} {Fan, S.},\
  }\bibfield  {title} {\bibinfo {title} {\emph {{Synthetic dimension in
  photonics}}},\ }\href {\doibase http://dx.doi.org/10.1364/OPTICA.5.001396}
  {\bibfield  {journal} {\bibinfo  {journal} {Optica}\ }\textbf {\bibinfo
  {volume} {5}},\ \bibinfo {pages} {1396} (\bibinfo {year} {2018})}\BibitemShut
  {NoStop}%
\bibitem [{\citenamefont {D'Errico}\ \emph {et~al.}(2021)\citenamefont
  {D'Errico}, \citenamefont {Barboza}, \citenamefont {Tudor}, \citenamefont
  {Dauphin}, \citenamefont {Massignan}, \citenamefont {Marrucci},\ and\
  \citenamefont {Cardano}}]{DErrico2021}%
  \BibitemOpen
  \bibfield  {author} {\bibinfo {author} {D'Errico, A.}, \bibinfo {author}
  {Barboza, R.}, \bibinfo {author} {Tudor, R.}, \bibinfo {author} {Dauphin,
  A.}, \bibinfo {author} {Massignan, P.}, \bibinfo {author} {Marrucci, L.}\
  and\ \bibinfo {author} {Cardano, F.},\ }\bibfield  {title} {\bibinfo {title}
  {\emph {{Bloch--Landau--Zener dynamics induced by a synthetic field in a
  photonic quantum walk}}},\ }\href {\doibase 10.1063/5.0037327} {\bibfield
  {journal} {\bibinfo  {journal} {APL Photonics}\ }\textbf {\bibinfo {volume}
  {6}},\ \bibinfo {pages} {020802} (\bibinfo {year} {2021})}\BibitemShut
  {NoStop}%
\bibitem [{\citenamefont {Piccardo}\ \emph {et~al.}(2022)\citenamefont
  {Piccardo}, \citenamefont {Ginis}, \citenamefont {Forbes}, \citenamefont
  {Mahler}, \citenamefont {Friesem}, \citenamefont {Davidson}, \citenamefont
  {Ren}, \citenamefont {Dorrah}, \citenamefont {Capasso}, \citenamefont
  {Dullo}, \citenamefont {Ahluwalia}, \citenamefont {Ambrosio}, \citenamefont
  {Gigan}, \citenamefont {Treps}, \citenamefont {Hiekkam{\"{a}}ki},
  \citenamefont {Fickler}, \citenamefont {Kues}, \citenamefont {Moss},
  \citenamefont {Morandotti}, \citenamefont {Riemensberger}, \citenamefont
  {Kippenberg}, \citenamefont {Faist}, \citenamefont {Scalari}, \citenamefont
  {Picqu{\'{e}}}, \citenamefont {H{\"{a}}nsch}, \citenamefont {Cerullo},
  \citenamefont {Manzoni}, \citenamefont {Lugiato}, \citenamefont {Brambilla},
  \citenamefont {Columbo}, \citenamefont {Gatti}, \citenamefont {Prati},
  \citenamefont {Shiri}, \citenamefont {Abouraddy}, \citenamefont {Al{\`{u}}},
  \citenamefont {Galiffi}, \citenamefont {Pendry},\ and\ \citenamefont
  {Huidobro}}]{Piccardo2022}%
  \BibitemOpen
  \bibfield  {author} {\bibinfo {author} {Piccardo, M.} \emph {et~al.},\
  }\bibfield  {title} {\bibinfo {title} {\emph {{Roadmap on multimode light
  shaping}}},\ }\href {\doibase 10.1088/2040-8986/ac3a9d} {\bibfield  {journal}
  {\bibinfo  {journal} {J. Opt.}\ }\textbf {\bibinfo {volume} {24}},\ \bibinfo
  {pages} {013001} (\bibinfo {year} {2022})}\BibitemShut {NoStop}%
\bibitem [{\citenamefont {Simon}\ and\ \citenamefont
  {Mukunda}(1990)}]{Simon1990}%
  \BibitemOpen
  \bibfield  {author} {\bibinfo {author} {Simon, R.}\ and\ \bibinfo {author}
  {Mukunda, N.},\ }\bibfield  {title} {\bibinfo {title} {\emph {{Minimal
  three-component SU(2) gadget for polarization optics}}},\ }\href {\doibase
  https://dx.doi.org/10.1016/0375-9601(90)90732-4} {\bibfield  {journal}
  {\bibinfo  {journal} {Phys. Lett. A}\ }\textbf {\bibinfo {volume} {143}},\
  \bibinfo {pages} {165} (\bibinfo {year} {1990})}\BibitemShut {NoStop}%
\bibitem [{\citenamefont {Devlin}\ \emph
  {et~al.}(2017{\natexlab{a}})\citenamefont {Devlin}, \citenamefont {Ambrosio},
  \citenamefont {Rubin}, \citenamefont {Mueller},\ and\ \citenamefont
  {Capasso}}]{Devlin2017}%
  \BibitemOpen
  \bibfield  {author} {\bibinfo {author} {Devlin, R.~C.}, \bibinfo {author}
  {Ambrosio, A.}, \bibinfo {author} {Rubin, N.~A.}, \bibinfo {author} {Mueller,
  J. P.~B.}\ and\ \bibinfo {author} {Capasso, F.},\ }\bibfield  {title}
  {\bibinfo {title} {\emph {Arbitrary spin-to-orbital angular momentum
  conversion of light}},\ }\href {\doibase 10.1126/science.aao5392} {\bibfield
  {journal} {\bibinfo  {journal} {Science}\ }\textbf {\bibinfo {volume}
  {358}},\ \bibinfo {pages} {896} (\bibinfo {year}
  {2017}{\natexlab{a}})}\BibitemShut {NoStop}%
\bibitem [{\citenamefont {Devlin}\ \emph
  {et~al.}(2017{\natexlab{b}})\citenamefont {Devlin}, \citenamefont {Ambrosio},
  \citenamefont {Wintz}, \citenamefont {Oscurato}, \citenamefont {Zhu},
  \citenamefont {Khorasaninejad}, \citenamefont {Oh}, \citenamefont
  {Maddalena},\ and\ \citenamefont {Capasso}}]{Devlin2017b}%
  \BibitemOpen
  \bibfield  {author} {\bibinfo {author} {Devlin, R.~C.}, \bibinfo {author}
  {Ambrosio, A.}, \bibinfo {author} {Wintz, D.}, \bibinfo {author} {Oscurato,
  S.~L.}, \bibinfo {author} {Zhu, A.~Y.}, \bibinfo {author} {Khorasaninejad,
  M.}, \bibinfo {author} {Oh, J.}, \bibinfo {author} {Maddalena, P.}\ and\
  \bibinfo {author} {Capasso, F.},\ }\bibfield  {title} {\bibinfo {title}
  {\emph {{Spin-to-orbital angular momentum conversion in dielectric
  metasurfaces}}},\ }\href {\doibase 10.1364/OE.25.000377} {\bibfield
  {journal} {\bibinfo  {journal} {Opt. Express}\ }\textbf {\bibinfo {volume}
  {25}},\ \bibinfo {pages} {377} (\bibinfo {year}
  {2017}{\natexlab{b}})}\BibitemShut {NoStop}%
\bibitem [{\citenamefont {Dorrah}\ and\ \citenamefont
  {Capasso}(2022)}]{Dorrah2022}%
  \BibitemOpen
  \bibfield  {author} {\bibinfo {author} {Dorrah, A.~H.}\ and\ \bibinfo
  {author} {Capasso, F.},\ }\bibfield  {title} {\bibinfo {title} {\emph
  {{Tunable structured light with flat optics}}},\ }\href
  {https://www.science.org/doi/10.1126/science.abi6860} {\bibfield  {journal}
  {\bibinfo  {journal} {Science}\ }\textbf {\bibinfo {volume} {376}},\ \bibinfo
  {pages} {367} (\bibinfo {year} {2022})}\BibitemShut {NoStop}%
\bibitem [{\citenamefont {Flamini}\ \emph {et~al.}(2018)\citenamefont
  {Flamini}, \citenamefont {Spagnolo},\ and\ \citenamefont
  {Sciarrino}}]{Flamini2018}%
  \BibitemOpen
  \bibfield  {author} {\bibinfo {author} {Flamini, F.}, \bibinfo {author}
  {Spagnolo, N.}\ and\ \bibinfo {author} {Sciarrino, F.},\ }\bibfield  {title}
  {\bibinfo {title} {\emph {{Photonic quantum information processing: a
  review}}},\ }\href {\doibase 10.1088/1361-6633/aad5b2} {\bibfield  {journal}
  {\bibinfo  {journal} {Reports Prog. Phys.}\ }\textbf {\bibinfo {volume}
  {82}},\ \bibinfo {pages} {016001} (\bibinfo {year} {2018})}\BibitemShut
  {NoStop}%
\bibitem [{\citenamefont {Regensburger}\ \emph {et~al.}(2012)\citenamefont
  {Regensburger}, \citenamefont {Bersch}, \citenamefont {Miri}, \citenamefont
  {Onishchukov}, \citenamefont {Christodoulides},\ and\ \citenamefont
  {Peschel}}]{Regensburger2012}%
  \BibitemOpen
  \bibfield  {author} {\bibinfo {author} {Regensburger, A.}, \bibinfo {author}
  {Bersch, C.}, \bibinfo {author} {Miri, M.-A.}, \bibinfo {author}
  {Onishchukov, G.}, \bibinfo {author} {Christodoulides, D.~N.}\ and\ \bibinfo
  {author} {Peschel, U.},\ }\bibfield  {title} {\bibinfo {title} {\emph
  {{Parity--time synthetic photonic lattices}}},\ }\href {\doibase
  https://doi.org/10.1038/nature11298} {\bibfield  {journal} {\bibinfo
  {journal} {Nature}\ }\textbf {\bibinfo {volume} {488}},\ \bibinfo {pages}
  {167} (\bibinfo {year} {2012})}\BibitemShut {NoStop}%
\bibitem [{\citenamefont {Wu}\ \emph {et~al.}(2020)\citenamefont {Wu},
  \citenamefont {Izaac}, \citenamefont {Li}, \citenamefont {Wang},
  \citenamefont {Chen}, \citenamefont {Zhu}, \citenamefont {Wang},\ and\
  \citenamefont {Ma}}]{Wu2020}%
  \BibitemOpen
  \bibfield  {author} {\bibinfo {author} {Wu, T.}, \bibinfo {author} {Izaac,
  J.~A.}, \bibinfo {author} {Li, Z.-X.}, \bibinfo {author} {Wang, K.}, \bibinfo
  {author} {Chen, Z.-Z.}, \bibinfo {author} {Zhu, S.}, \bibinfo {author} {Wang,
  J.~B.}\ and\ \bibinfo {author} {Ma, X.-S.},\ }\bibfield  {title} {\bibinfo
  {title} {\emph {{Experimental Parity-Time Symmetric Quantum Walks for
  Centrality Ranking on Directed Graphs}}},\ }\href {\doibase
  https://dx.doi.org/10.1103/PhysRevLett.125.240501} {\bibfield  {journal}
  {\bibinfo  {journal} {Phys. Rev. Lett.}\ }\textbf {\bibinfo {volume} {125}},\
  \bibinfo {pages} {240501} (\bibinfo {year} {2020})}\BibitemShut {NoStop}%
\bibitem [{\citenamefont {Nielsen}\ and\ \citenamefont
  {Chuang}(2010)}]{nielsen_chuang_2010}%
  \BibitemOpen
  \bibfield  {author} {\bibinfo {author} {Nielsen, M.~A.}\ and\ \bibinfo
  {author} {Chuang, I.~L.},\ }\href {\doibase 10.1017/CBO9780511976667} {\emph
  {\bibinfo {title} {Quantum Computation and Quantum Information: 10th
  Anniversary Edition}}}\ (\bibinfo  {publisher} {Cambridge University Press},\
  \bibinfo {year} {2010})\BibitemShut {NoStop}%
\bibitem [{\citenamefont {Vieira}\ \emph {et~al.}(2013)\citenamefont {Vieira},
  \citenamefont {Amorim},\ and\ \citenamefont {Rigolin}}]{Vieira2013}%
  \BibitemOpen
  \bibfield  {author} {\bibinfo {author} {Vieira, R.}, \bibinfo {author}
  {Amorim, E. P.~M.}\ and\ \bibinfo {author} {Rigolin, G.},\ }\bibfield
  {title} {\bibinfo {title} {\emph {{Dynamically Disordered Quantum Walk as a
  Maximal Entanglement Generator}}},\ }\href {\doibase
  http://dx.doi.org/10.1103/PhysRevLett.111.180503} {\bibfield  {journal}
  {\bibinfo  {journal} {Phys. Rev. Lett.}\ }\textbf {\bibinfo {volume} {111}},\
  \bibinfo {pages} {180503} (\bibinfo {year} {2013})}\BibitemShut {NoStop}%
\bibitem [{\citenamefont {Wang}\ \emph {et~al.}(2018)\citenamefont {Wang},
  \citenamefont {Xu}, \citenamefont {Pan}, \citenamefont {Sun}, \citenamefont
  {Xu}, \citenamefont {Chen}, \citenamefont {Han}, \citenamefont {Li},\ and\
  \citenamefont {Guo}}]{Wang2018}%
  \BibitemOpen
  \bibfield  {author} {\bibinfo {author} {Wang, Q.-Q.}, \bibinfo {author} {Xu,
  X.-Y.}, \bibinfo {author} {Pan, W.-W.}, \bibinfo {author} {Sun, K.}, \bibinfo
  {author} {Xu, J.-S.}, \bibinfo {author} {Chen, G.}, \bibinfo {author} {Han,
  Y.-J.}, \bibinfo {author} {Li, C.-F.}\ and\ \bibinfo {author} {Guo, G.-C.},\
  }\bibfield  {title} {\bibinfo {title} {\emph {{Dynamic-disorder-induced
  enhancement of entanglement in photonic quantum walks}}},\ }\href {\doibase
  https://doi.org/10.1364/OPTICA.5.001136} {\bibfield  {journal} {\bibinfo
  {journal} {Optica}\ }\textbf {\bibinfo {volume} {5}},\ \bibinfo {pages}
  {1136} (\bibinfo {year} {2018})}\BibitemShut {NoStop}%
\bibitem [{\citenamefont {James}\ \emph {et~al.}(2001)\citenamefont {James},
  \citenamefont {Kwiat}, \citenamefont {Munro},\ and\ \citenamefont
  {White}}]{Jame01_PRA}%
  \BibitemOpen
  \bibfield  {author} {\bibinfo {author} {James, D. F.~V.}, \bibinfo {author}
  {Kwiat, P.~G.}, \bibinfo {author} {Munro, W.~J.}\ and\ \bibinfo {author}
  {White, A.~G.},\ }\bibfield  {title} {\bibinfo {title} {\emph {{Measurement
  of qubits}}},\ }\href {\doibase 10.1103/PhysRevA.64.052312} {\bibfield
  {journal} {\bibinfo  {journal} {Phys. Rev. A}\ }\textbf {\bibinfo {volume}
  {64}},\ \bibinfo {pages} {052312} (\bibinfo {year} {2001})}\BibitemShut
  {NoStop}%
\bibitem [{\citenamefont {Esposito}\ \emph {et~al.}(2022)\citenamefont
  {Esposito}, \citenamefont {Barros}, \citenamefont {{Dur{\'{a}}n
  Hern{\'{a}}ndez}}, \citenamefont {Carvacho}, \citenamefont {{Di Colandrea}},
  \citenamefont {Barboza}, \citenamefont {Cardano}, \citenamefont {Spagnolo},
  \citenamefont {Marrucci},\ and\ \citenamefont {Sciarrino}}]{Esposito2022}%
  \BibitemOpen
  \bibfield  {author} {\bibinfo {author} {Esposito, C.}, \bibinfo {author}
  {Barros, M.~R.}, \bibinfo {author} {{Dur{\'{a}}n Hern{\'{a}}ndez}, A.},
  \bibinfo {author} {Carvacho, G.}, \bibinfo {author} {{Di Colandrea}, F.},
  \bibinfo {author} {Barboza, R.}, \bibinfo {author} {Cardano, F.}, \bibinfo
  {author} {Spagnolo, N.}, \bibinfo {author} {Marrucci, L.}\ and\ \bibinfo
  {author} {Sciarrino, F.},\ }\bibfield  {title} {\bibinfo {title} {\emph
  {{Quantum walks of two correlated photons in a 2D synthetic lattice}}},\
  }\href {\doibase 10.1038/s41534-022-00544-0} {\bibfield  {journal} {\bibinfo
  {journal} {npj Quantum Inf.}\ }\textbf {\bibinfo {volume} {8}},\ \bibinfo
  {pages} {34} (\bibinfo {year} {2022})}\BibitemShut {NoStop}%
\end{thebibliography}
\end{document}